\documentclass{aa}  

\usepackage{caption}
\usepackage{subcaption}

\usepackage{natbib}
\usepackage[colorlinks,citecolor=blue,urlcolor=blue,bookmarks=false,hypertexnames=true]{hyperref} 
\usepackage{graphicx}
\usepackage{tikz}

\begin{document} 

   \title{Quiet Sun flux rope formation via incomplete Taylor relaxation}
   
  % \subtitle{Ordering chaos in a fully stratified 3D MHD simulation}
   % Redistribution of magnetic helicity in a convection-driven flux rope

   \author{Rebecca A. Robinson
          \inst{1}$^,$\inst{2}
         \and
         Guillaume Aulanier \inst{3}$^,$\inst{1}
          \and
          Mats Carlsson \inst{1}$^,$\inst{2}
          }

        %   \and
        %   C. Ptolemy\inst{2}\fnmsep\thanks{Just to show the usage
        %   of the elements in the author field}
        %   }

   \institute{Rosseland Centre for Solar Physics (RoCS), University of Oslo, P.O. Box 1029, Blindern, NO-0315 Oslo, Norway \\
            %   \email{rebecrob@astro.uio.no}
              \and
             Institute of Theoretical Astrophysics, University of Oslo, P.O. Box 1029, Blindern, NO-0315 Oslo, Norway\\
             \and 
             Sorbonne Universit\'e, \'Ecole Polytechnique, Institut Polytechnique de Paris, Observatoire de Paris - PSL, CNRS, Laboratoire de physique des plasmas (LPP), 4 place Jussieu, F-75005 Paris, France \\
            %  \email{c.ptolemy@hipparch.uheaven.space}
            %  \thanks{The university of heaven temporarily does not
            %          accept e-mails}
             }

   \date{}

% % \abstract{}{}{}{}{} 
% % 5 {} token are mandatory
 
  \abstract
%   context heading (optional)
  {Low-altitude nanoflares are among the candidates for atmospheric heating in the quiet Sun's corona. Low-altitude twisted magnetic fields may be involved in such events, as they are in larger flares. But for nanoflares, the exact role, topology, and formation mechanisms of these twisted fields remain to be studied.}
  {In this paper, we investigate the formation and evolution of a preflare flux rope in a fully stratified, 3D magnetohydrodynamics (MHD) simulation of the quiet Sun using the \emph{Bifrost} code. This study focuses on the time period before the rope eventually reconnects with an overlying field, resulting in a nanoflare-scale energy on the order of $10^{17}$ J. One puzzle is that this modeled flux rope does not form by any of the mechanisms usually at work in larger flares, such as flux emergence, flux cancellation, or tether-cutting reconnection.}
   {Using Lagrangian markers to trace representative field lines, we follow the spatiotemporal evolution of the flux rope. By focusing on current volumes (which we call current sheets) between these lines, we identify flux bundles and associated reconnecting field line pairs. We also analyze the time-varying distribution function for the force-free parameter as the flux rope relaxes. Lastly, we compare different seeding methods for tracing magnetic field lines, and discuss their relevance to the analysis.}
  {We show that the modeled flux rope is gradually built from the coalescence of numerous current-carrying flux tubes. This occurs through a series of component reconnections that are continuously driven by the complex flows in the underlying convection zone. These reconnections lead to an inverse cascade of helicity from small scales to larger scales. We also find that the system attempts to relax toward a linear force-free field, but that the convective drivers and the nanoflare event prevent full Taylor relaxation.}  
  {Using a self-consistently driven simulation of a nanoflare event, we show for the first time an inverse helicity cascade tending toward a Taylor relaxation in the Sun's corona, resulting in a well-ordered flux rope that later reconnects with surrounding fields. This provides context clues toward understanding the buildup of nanoflare events in the quiet Sun through incomplete Taylor relaxations, when no relevant flux emergence or cancellation is observed.}

  \maketitle
%
%-------------------------------------------------------------------

\section{Introduction}
% % What is known?
The majority of the Sun is covered by a network of relatively weak magnetic features that together comprise the quiet Sun, and it is now standard practice to consider quiet Sun magnetism a reliable source of atmospheric energy \citep{2011SSRv..159..263H, 2019LRSP...16....1B}. Although small-scale energetic events such as nanoflares have been observed and simulated \citep[e.g.,][]{1988ApJ...330..474P,2013ApJ...770L...1T,2014Sci...346B.315T,2018A&A...620L...5B}, it can be a challenge to understand the magnetic field evolution that leads to small-scale energy release. Additionally, despite a wealth of multiwavelength observations and simulations of varying degrees of complexity, relatively little is known about how magnetic topologies at small scales form and evolve.  

The mechanisms behind the formation of flux ropes have been studied extensively in simulations \citep[e.g.,][]{1999ApJ...515L..81A,1999ApJ...518L..57A,1999A&A...351..707T,2009ApJ...697.1529F,2010ApJ...708..314A,2016A&A...587A.125P,2016A&A...591A..16P} and observations \citep[e.g.,][]{2016ApJ...818..148L,2019FrASS...6...63J}. These studies span a variety of different energies from simple flux emergence, to noneruptive events, to X-class flares. Generally, simulations can be used to focus on the buildup and evolution of flux ropes under given conditions by following their formation in time. On the other hand, reconstructing an observed flux rope depends on nonlinear force-free extrapolations of an observed magnetogram, which are then compared to multiwavelength observations of the same region \citep[e.g.,][]{2011ApJ...732L..25C,2014ApJ...789L..35C,2015ApJ...808L..15S}. From either direction, understanding the formation of magnetic flux ropes is integral to understanding the magnetic conditions that lead to reconnection events and energy release in large eruptive flares \citep{2014IAUS..300..184A} as well as small-scale eruptive events. Frequently observed and well-known flux rope evolution processes include flux emergence \citep{2004A&A...426.1047A}, flux cancellation \citep{1989ApJ...343..971V}, and tether-cutting reconnection \citep{1992LNP...399...69M}.

Another process by which flux ropes form and evolve is the so-called inverse cascade of helicity, wherein the helicity of a flux system is redistributed from smaller bundles to a larger twisted rope via a series of small-scale reconnections in the atmosphere \citep{1975JFM....68..769F,2019E&SS....6..351P}. This can happen when there is cross-helicity between separate, adjacent flux systems, or when narrow twisted flux tubes within the system contain self-helicities and reconnect with one another. This process has been theoretically established and modeled for plane-parallel loops-in-a-box \citep[e.g.,][]{1999GMS...111..187A,2007A&A...473..615W,2010PhRvL.105h5002Y,2015ApJ...805...61Z,2017ApJ...835...85K,2019ApJ...883..148R}. It has been shown that the redistribution of helicity from a collection of small flux bundles to a larger-scale helical system is dependent on whether or not the conditions for reconnection are met as a result of rotating photospheric drivers. With that, two fundamental unknowns arise. The first concerns whether or not this process can generate a long, coherent flux rope from an incoherent collection of field lines in the corona, thus establishing an additional mechanism for flux rope formation before flaring. The second unknown concerns the final distribution of currents in the flux rope in that case.

As flux systems form and evolve, they relax to some final helical state over time which, according to Taylor's theory, should approach a linear force-free field \citep{1974PhRvL..33.1139T,1999PPCF...41B.167B,Pariat2020,Yeates2020}. However, this is not something that is seen to occur in the observed corona or in simpler coronal models, as demonstrated in idealized, line-tied simulations with well-ordered magnetic fields \citep{1999GMS...111..187A,1999PPCF...41A.779A}. This is because the topological constraints (that is, line tying) are such that the field cannot fully relax to a linear force-free field. Specifically, such topological constraints prevent the ubiquitous redistribution of currents and helicity, and limit the possibility of reconnection to only certain locations. With that, it is unclear whether there are still cases in which solar magnetic fields tend to relax according to Taylor's theory, or whether this process is not possible at all under solar conditions.

% What is our response?
This work follows our previous work in \citet{2022A&A...668A.177R}, in which our convection-driven simulation using the \textit{Bifrost} code revealed the ordering of a magnetic arcade and a  flux rope. These features eventually reconnected with an overlying horizontal field, resulting in a coronal heating event with plasma temperatures up to 1.47 MK and an integrated Joule heating energy of 5.4 $\times 10^{17}$ J. The longevity of the arcade and flux rope is related to their long-lasting photospheric roots, but the mechanisms behind their formation and atmospheric ordering were unclear. We observed a gradual development of large-scale and relatively uniform twist from a system of tangled, low-lying loops and suggested that this field ordering may have been a result of inverse helicity cascade, but left the suggestion open. In this paper, we present evidence that the ordering of the flux rope in particular is dictated by several small-scale reconnection events, redistributing the magnetic helicity from small scales to larger scales until the rope is organized. In addition, we show for the first time in a stratified simulation that the flux rope tends to evolve toward a linear force-free field, hence displaying evidence of an incomplete Taylor relaxation in the corona.  

% Reader guide
This paper is the second in a series; the first discussed the magnetic geometry of a quiet Sun magnetic reconnection event, which falls within the nanoflare regime \citep{2022A&A...668A.177R}. This paper addresses the formation of the previously discussed magnetic flux rope before the onset of reconnection. A third paper will discuss the observational signatures of the simulated nanoflare event.

% %--------------------------------------------------------------------
\section{Methods}
\subsection{The \textit{Bifrost} simulation}
The parallel numerical code \textit{Bifrost} solves the MHD equations within the context of stellar atmospheres, from the upper convection zone to the corona. Its design and implementation is discussed in detail in \citet{2011A&A...531A.154G}. In addition to its ability to include multiple atmospheric layers which have vastly different physical conditions, \textit{Bifrost} includes several built-in modules for different physical assumptions, boundary conditions, test particles, and more. 

The time stepping recipe that we use is explicit and third-order, as given in \citet{hyman}. Radiative transfer in the upper photosphere and lower chromosphere employs multigroup opacities with four opacity bins \citep{1982A&A...107....1N} and scattering \citep{2000ApJ...536..465S}, and is implemented using a short characteristics scheme that follows \citet{2010A&A...517A..49H}. The energy budget of the upper chromosphere, transition region and corona are solved according to \citet{2012A&A...539A..39C}. The calculation of conduction along magnetic field lines, essential to the energy budget of the corona, follows the recipe in \citet{2017ApJ...834...10R}.

The simulation used for this study is the same one described in \citet{2022A&A...668A.177R}, which is a \textit{Bifrost} simulation of the quiet Sun. The complete history of this simulation is summarized in \citet{2022A&A...668A.177R}, but we repeat here that our segment of interest began with an initially balanced vertical magnetic field which evolved in time via self-consistent convective drivers. Our $512^3$ grid ranges from 2.5 Mm beneath the average $\tau_{500} = 1$ surface (our definition of z = 0) and extends to 8 Mm above it. Our vertical spacing is nonuniform; it begins at the lower convective boundary with 30 km resolution, then gradually sharpens to 12-14 km between z = 0 and z = 4 Mm, then eventually increases to a coarser 70.5 km at the upper coronal boundary. We note that in \textit{Bifrost}, our vertical coordinate refers to depth rather than height and therefore increases downward from the coronal boundary to the convective boundary. All vertical vector quantities are aligned with this geometry, and all of our 3D renderings reflect this standard. 

The horizontal extent of this simulation is 12 Mm in both horizontal dimensions with a uniform horizontal resolution of 23 km, and we employ periodic boundary conditions over the horizontal boundaries. Periodic boundary conditions allow the magnetic field lines to cross the horizontal boundaries and reenter on the other side, with some of them finding photospheric roots that would have been impossible with open horizontal boundaries. The lower convective boundary is open and allows flows at the prescribed entropy such that inflows maintain an effective temperature of $\approx 5780$ K, and the upper coronal boundary is open. This is a hyper-diffusion run, which prevents the formation of current volumes (as we call current sheets) that are smaller than the grid resolution, as well as the collapse of magnetic flux bundles at the numerical resolution  \citep{1995NordlundGalsgaard,2011A&A...531A.154G}. We consider hydrogen to be in local thermodynamic equilibrium (LTE) in this simulation.

\subsection{Visualizing magnetic fields in time and space}
In \citet{2022A&A...668A.177R}, we addressed the onset of a nanoflare-scale reconnection event that was likely a result of several small-scale reconnection events, which had ordered the field into two prominent magnetic features. Demonstrating those small-scale reconnections is possible but not straightforward in a stratified simulation with convection-driven photospheric flux concentrations, and a dynamic chromosphere on top of that. To find small-scale reconnection events, we must be able to seed magnetic field tracings using specific and unambiguous methods.

The Visualization and Analysis Platform for Ocean, Atmosphere, and Solar Researchers (VAPOR) software \citep{2019Atmos..10..488L,vapor} is a useful post-processing tool for seeding magnetic field tracings in a highly specified manner. It is possible to seed a magnetic field tracing within any grid-aligned rectangular volume, and one may choose a random distribution of seeds with or without a bias toward either large or small values of a specified variable. In this way, we were able to specify seeding regions on and around specific current sheets within our simulation box, and therefore explore the associated fields traced from those regions via a Runge-Kutta 4 integration scheme.

These seeding methods are sufficient for inspecting a snapshot of interest within the simulation, but constructing a time series of magnetic field line evolution requires a different approach. Our simulation outputs one snapshot every ten solar seconds at a timestep on the order of $10^{-3}$ s, meaning the simulation updates thousands of times before writing an output file for further inspection. To construct a more reliable time series at the time resolution of the simulation, Lagrangian markers were injected into the simulation at t = 9\,669 s using a \textit{Bifrost} module called \texttt{corks} \citep{2018A&A...614A.110Z, 2022A&A...665A...6D,2022A&A...668A.177R}. 

\subsection{Using \texttt{corks} as seeds}\label{cx}
The \texttt{corks} module introduces Lagrangian markers at every gridpoint in the simulation, which are updated at the simulation timestep, following the ideal transport of the fluid. For this simulation, the number of corks in the simulation box remains consistent. Assuming that the magnetic field also follows the fluid except in diffusive regions, the location of each cork over time gives information about the evolution of each associated field line. By finding the locations of the nearest corks to field lines of interest, we could then construct a time series of those associated field lines by using the locations of the corks as seeds for past and future magnetic field line tracings. 

The features of interest in \citet{2022A&A...668A.177R} are a magnetic arcade and a weakly twisted flux tube, both of which eventually reconnect with an overlying, nearly antiparallel horizontal field in the corona. This paper focuses on the weakly twisted flux tube and its formation. Since we have corks that follow the flow of the fluid throughout the simulation, we then have the full history of the lines that become, are, or used to be a part of the flux rope.

In this study, we used corks as seeds for tracing the development of the flux rope, based on their proximity to randomly seeded field lines of interest at one given time stamp. First, we isolated the flux rope lines used in \citet{2022A&A...668A.177R} and looked for thin current sheets within a slice orthogonal to the rope. Based on those thin current sheets, we isolated an overall flux rope by randomly seeding field lines in and through the orthogonal slice. Then, to construct a time series of the evolution of the flux rope, we found the nearest cork to each line and used the cork identification to follow each line forward and backward in time. Without the use of corks as Lagrangian markers, it would not be possible to follow consistent field lines throughout the simulation. 

\section{Results}
\subsection{Tracing the relevant flux system}
As discussed in \citet{2022A&A...668A.177R} and in the above section, the seeds chosen to trace the magnetic field necessarily determine the resulting field tracing. This means that incorrectly seeding a magnetic field would cause us to miss relevant magnetic structures. In an exercise detailed in Appendix \ref{appendix:a} but motivated here, we aim to demonstrate that incorrectly seeding magnetic fields leads to missing important information, but that this can be avoided if we are careful and consistent with our seeding methods.

In our simulation, we see the formation and subsequent reconnection of several magnetic features, including a flux rope. As we discuss in this paper, this flux rope is built up from smaller flux systems which coalesce eventually into a larger flux rope, which later reconnects with an overlying horizontal field in the corona. In order to accurately construct the history of this flux rope, we have meticulously chosen relevant magnetic field lines based on their locations within the flux system at the time the flux rope has already formed. Then, we used Lagrangian markers to reconstruct a time series of the flux rope formation from before the flux rope forms to the time it has formed and has begun to reconnect.

To demonstrate that we have indeed isolated the relevant flux system using the correct seeding and tracing methods, we have conducted a comparison study of three different methods. In this comparison study, we explored the consequences of different seeding methods and determined that our chosen method allows us to reliably trace the relevant flux system. We also determined that this flux rope is indeed built up from tangled coronal lines and becomes coherent over time, as previously stated in \citet{2022A&A...668A.177R}. For the interested reader, a detailed summary of this comparison study is available in Appendix \ref{appendix:a}.

%%%%%%%%%%%%%%%%%%%%%%%%%%%%%%%%%%%%%%%%%%%%%%%%%%%%%%%%%%%
\begin{figure*}
\begin{subfigure}{.49\textwidth}
  \centering
  % include first image
  \includegraphics[width=\linewidth]{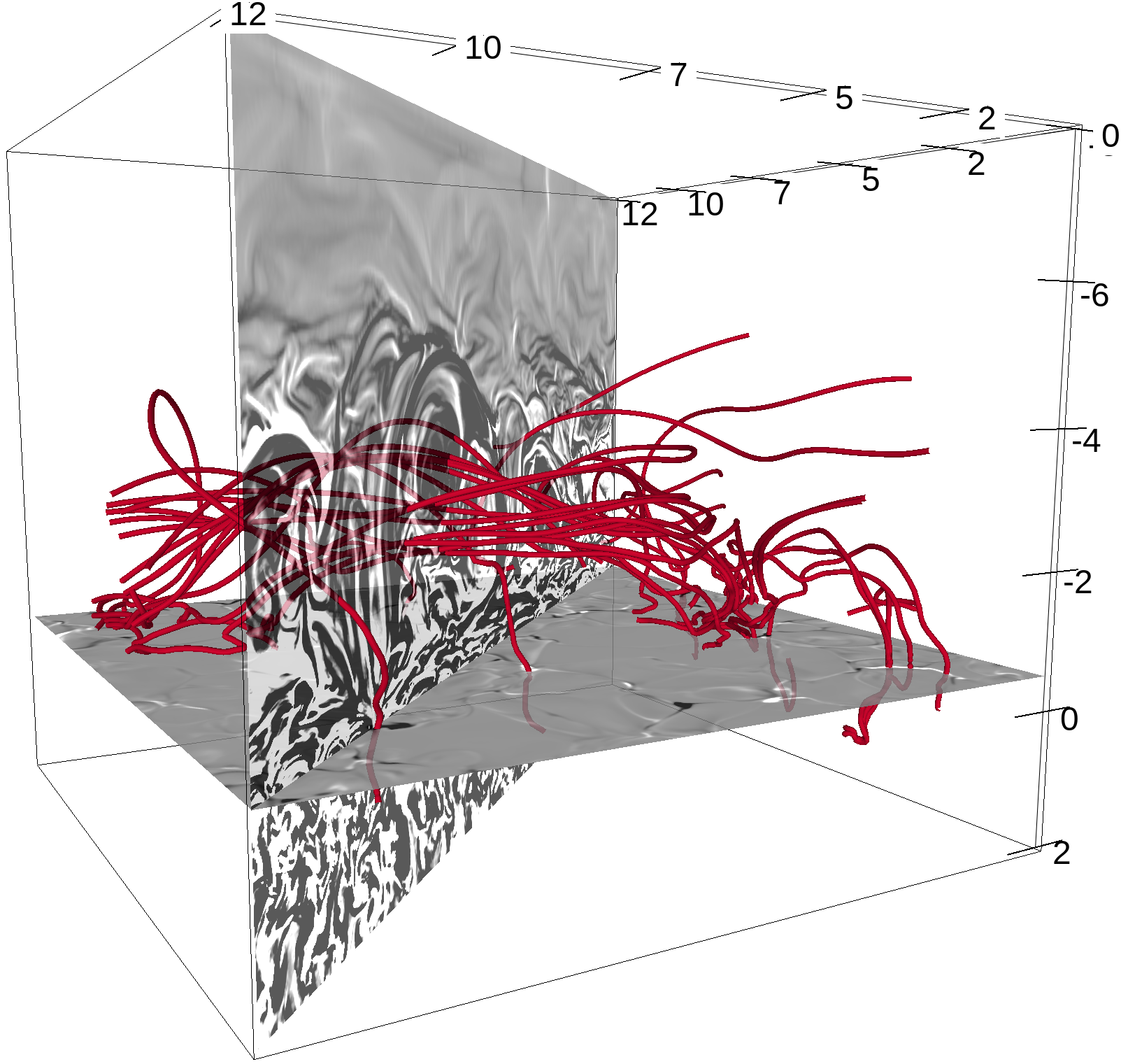}
\end{subfigure}
\begin{subfigure}{.48\textwidth}
  \centering
  % include second image
  \includegraphics[width=\linewidth]{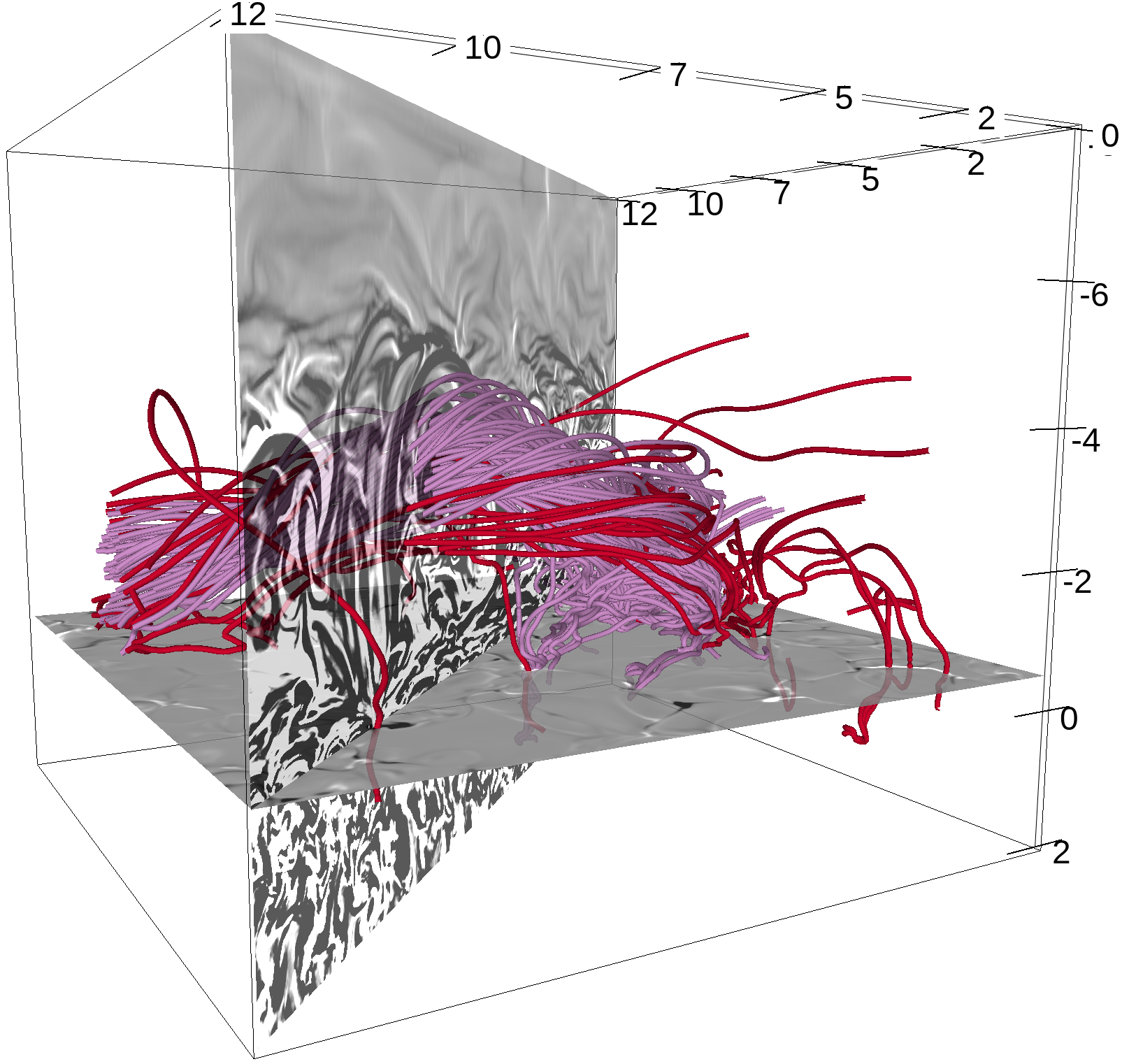}  
\end{subfigure}

\caption{Magnetic field lines at t = 11\,040 s belonging to a flux rope as analyzed in \citet{2022A&A...668A.177R}, with a bisecting plane colored by saturated $\alpha$ values (left). The right panel is the same as left panel, with the addition of lines corresponding to the wider flux rope that was first seeded by its proximity to the thin current sheets in the rope near the orthogonal slice, then seeded by corks that were geometrically closest to those lines at that time.}
\label{fig:flux_compare}
\end{figure*}

\subsection{General evolution of the overall flux rope}\label{flux}
% DISCUSS ALPHA HERE -- define equation & explain what the units are
In \citet{2022A&A...668A.177R}, we analyzed a set of field lines which form a small bundle that belongs to a wider flux rope. To find the extent of that flux rope when it has been formed but has not yet reconnected, we selected a slice orthogonal to the original selected flux rope lines and looked for thin current sheets within the rope. To find them, we use our functional definition of the force-free parameter $\alpha$ given in Equation \ref{eq:1}: 

\begin{equation}\label{eq:1}
    \alpha = \frac{\vec{\jmath}\cdot\vec{B}}{\mu_0B^2}
    = \frac{(\vec{\nabla}\times\vec{B})\cdot\vec{B}}{B^2}. 
\end{equation}

Here, $\vec{J}$, $\vec{B}$, and $\mu_0$ represent current density, magnetic field, and magnetic permeability as usual. This force-free parameter has units of inverse-distance, meaning largest absolute values of $\alpha$ pick out the thinnest current sheets, pointing to possible small-scale reconnection events. The force-free parameter is also directly proportional to twist in the overall flux system \citep{2006JPhA...39.8321B} as shown in Equation \ref{eq:2}:

\begin{equation}\label{eq:2}
    \frac{dT_w}{dl} =  \frac{(\vec{\nabla}\times\vec{B)}\cdot\vec{B}}{4 \pi B^2} \Rightarrow
        T_w = \int\frac{\alpha}{4\pi}dl \approx \frac{\alpha L}{4\pi}. 
\end{equation}

Here, $T_w$ refers to the number of turns in the flux system, $L$ is the length of the field line, and $dl$ represents the unit length. The last approximation is valid for any nearly force-free field, where $\alpha$ is constant along the field lines. 

The left panel of Figure \ref{fig:flux_compare} shows the selected flux rope lines from \citet{2022A&A...668A.177R} at t = 11\,040 s, a time during which the flux rope has developed twist. The vertical slice is a saturated 2D rendering of $\alpha$ in order to show the thinnest current sheets orthogonal to the flux rope; we note that this slice is reused as a reference cut throughout this paper. We chose this vertical cut at roughly the apex of the flux rope, and while it is evident that current sheets pervade the whole slice, the circular outline of the wider flux rope can be clearly seen. In addition, the cross-section provides a glimpse into the current sheets that exist within the flux rope at the given time. 

For comparison, the right panel of Figure \ref{fig:flux_compare} includes a rendering of the wider flux system. This rope was seeded first by a random distribution of seeds in and around the orthogonal slice, near and within the flux rope outline. Then, the nearest corks to those lines were chosen and used as seeds to create the rendering shown in the right panel of Figure \ref{fig:flux_compare}. Here, we see that the original flux rope (red lines) and the wider flux rope (pink lines) are members of the same flux system; having different positive roots depending on how the lines are seeded, but displaying consistent twist and handedness. 

\begin{figure*}
\begin{subfigure}{.49\textwidth}
  \centering
  \begin{tikzpicture}
  % include first image
  \node[anchor=south west,inner sep=0] (image) at (0,0)
  {\includegraphics[width=\linewidth]{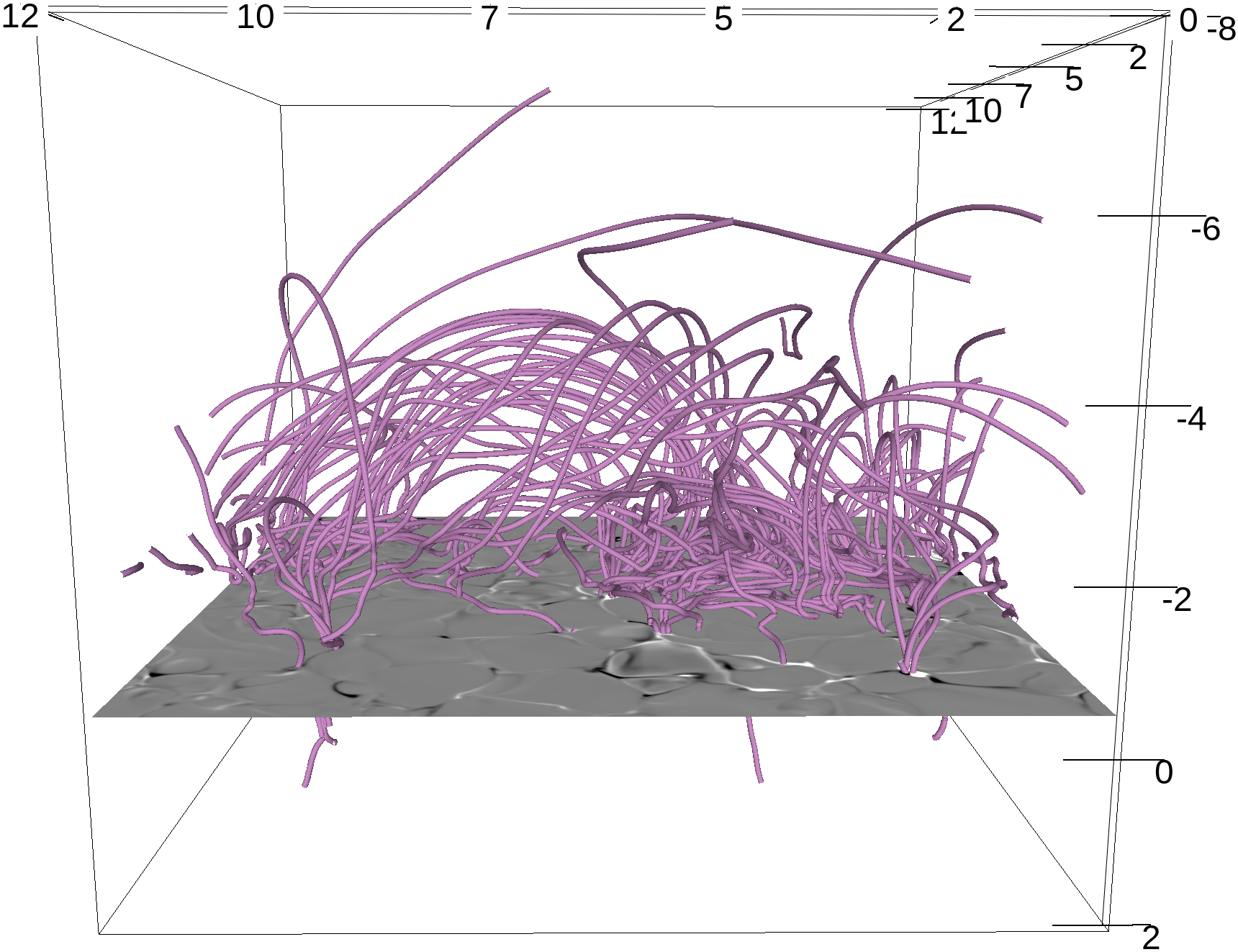}};
  \begin{scope}[x={(image.south east)},y={(image.north west)}]
    \node[black,fill=white,draw] at (0.3,0.9) {t* - 1371 (t = 9\,669 s)};
  \end{scope}
  \end{tikzpicture}
  % \subcaption[]{t* - 1371 (t = 9\,669 s)}
\end{subfigure}
\begin{subfigure}{.49\textwidth}
  \centering
    \begin{tikzpicture}
    \node[anchor=south west,inner sep=0] (image) at (0,0)
        {\includegraphics[width=\linewidth]{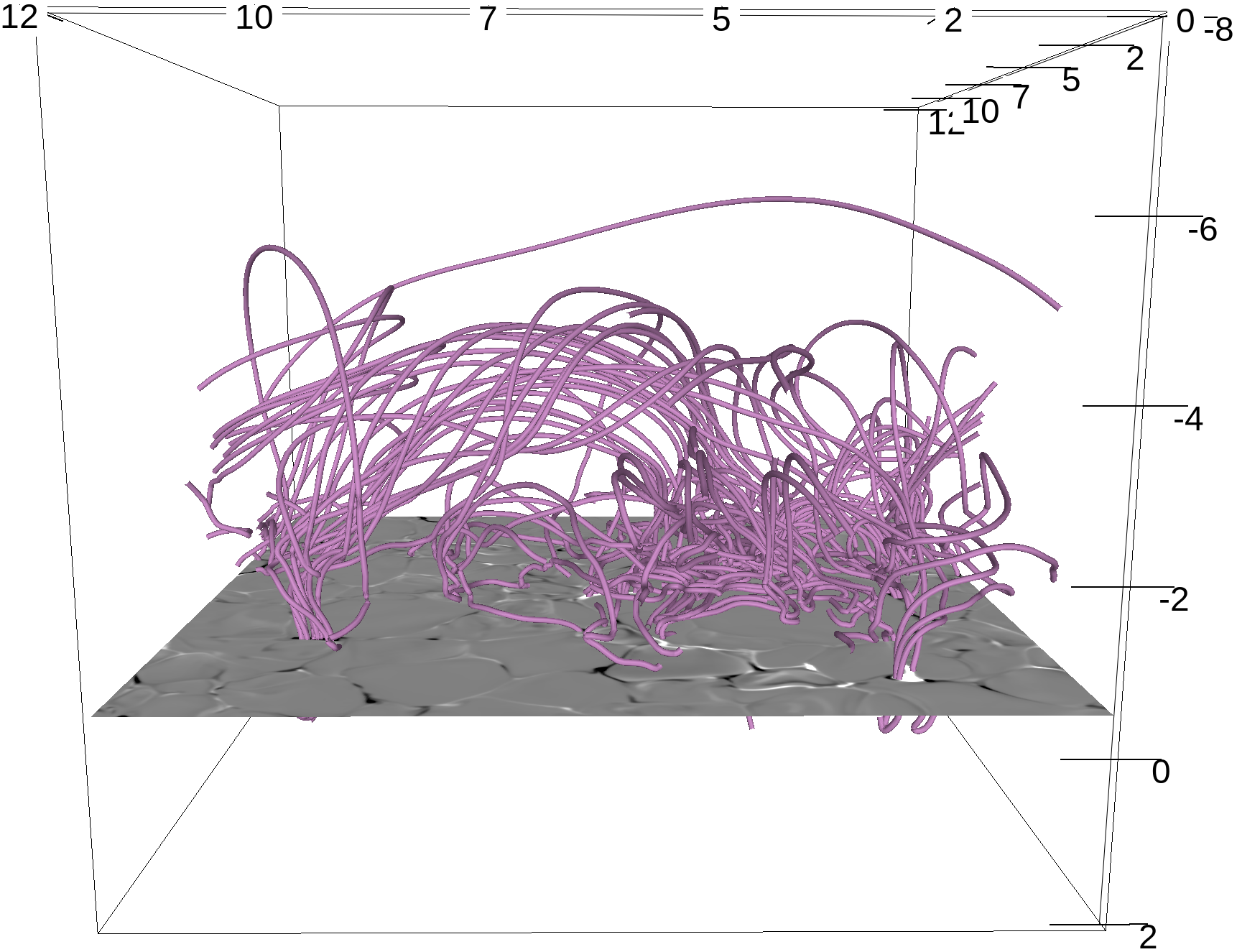}};
    \begin{scope}[x={(image.south east)},y={(image.north west)}]
    \node[black,fill=white,draw] at (0.3,0.9) {t* - 1201 (t = 9\,839 s)};
    \end{scope}
    \end{tikzpicture}

  % \subcaption[]{t* - 1201 (t = 9\,839 s)}
\end{subfigure}
\begin{subfigure}{.49\textwidth}
  \centering
    \begin{tikzpicture}
    \node[anchor=south west,inner sep=0] (image) at (0,0)
        {\includegraphics[width=\linewidth]{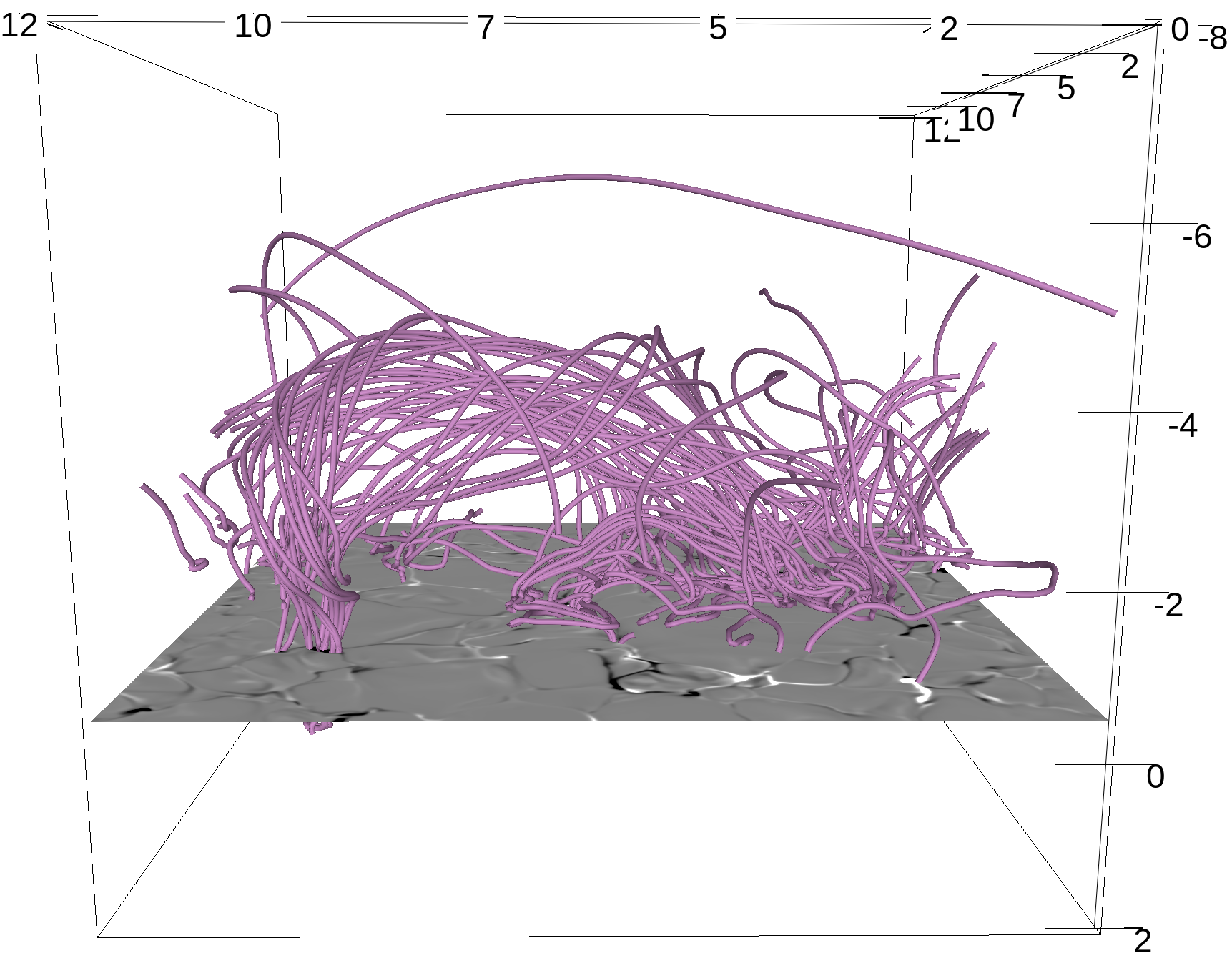}};
    \begin{scope}[x={(image.south east)},y={(image.north west)}]
    \node[black,fill=white,draw] at (0.3,0.9) {t* - 801 (t = 10\,239 s)};
    \end{scope}
    \end{tikzpicture}
  % \subcaption[]{t* - 801 (t = 10\,239 s)}
\end{subfigure}
\begin{subfigure}{.49\textwidth}
  \centering
    \begin{tikzpicture}
    \node[anchor=south west,inner sep=0] (image) at (0,0)
        {\includegraphics[width=\linewidth]{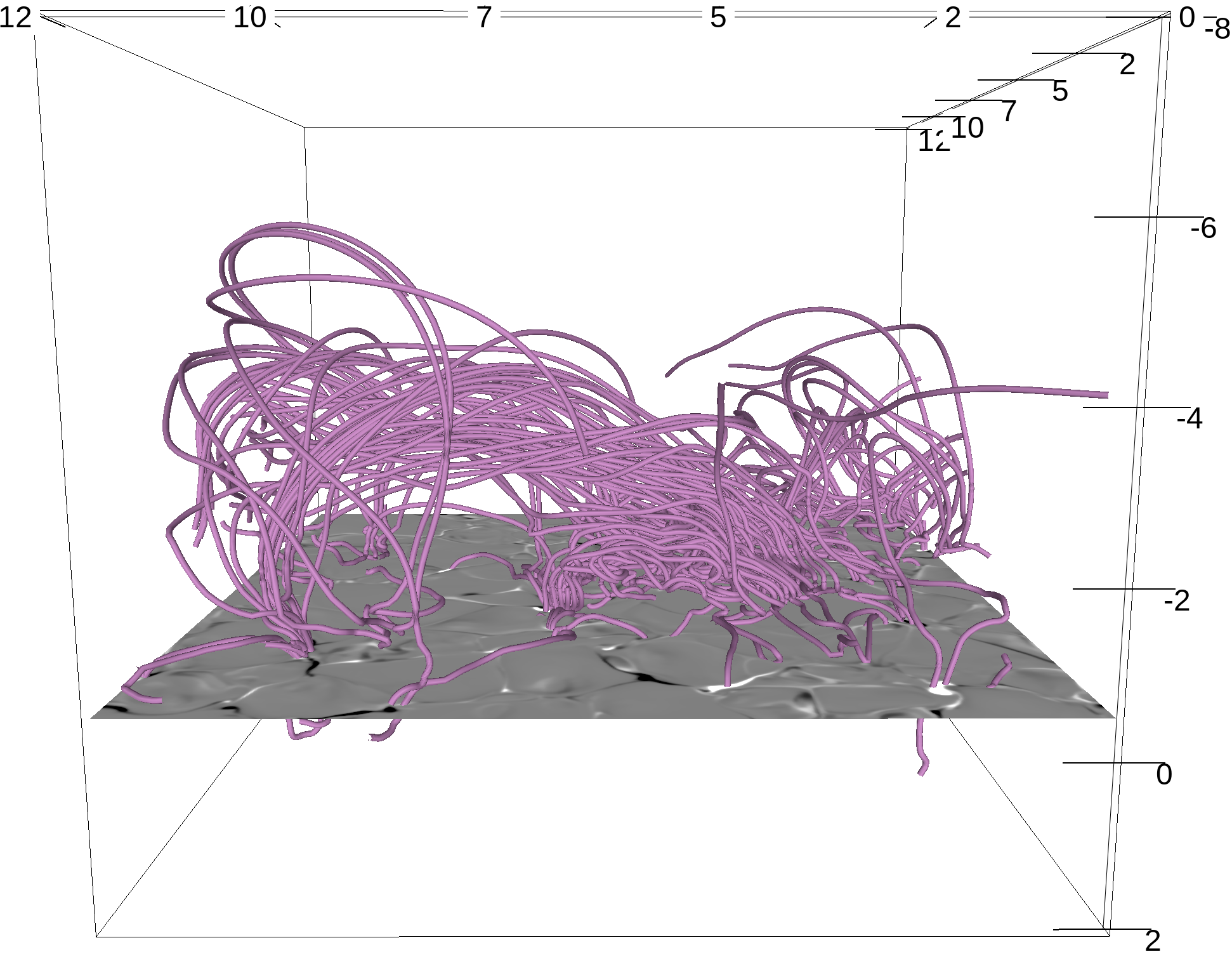}};
    \begin{scope}[x={(image.south east)},y={(image.north west)}]
    \node[black,fill=white,draw] at (0.3,0.9) {t* - 400 (t = 10\,640 s)};
    \end{scope}
    \end{tikzpicture} 
  % \subcaption[]{t* - 400 (t = 10\,640 s)}
\end{subfigure}
\begin{subfigure}{.49\textwidth}
  \centering
  % include first image
    \begin{tikzpicture}
    \node[anchor=south west,inner sep=0] (image) at (0,0)
        {\includegraphics[width=\linewidth]{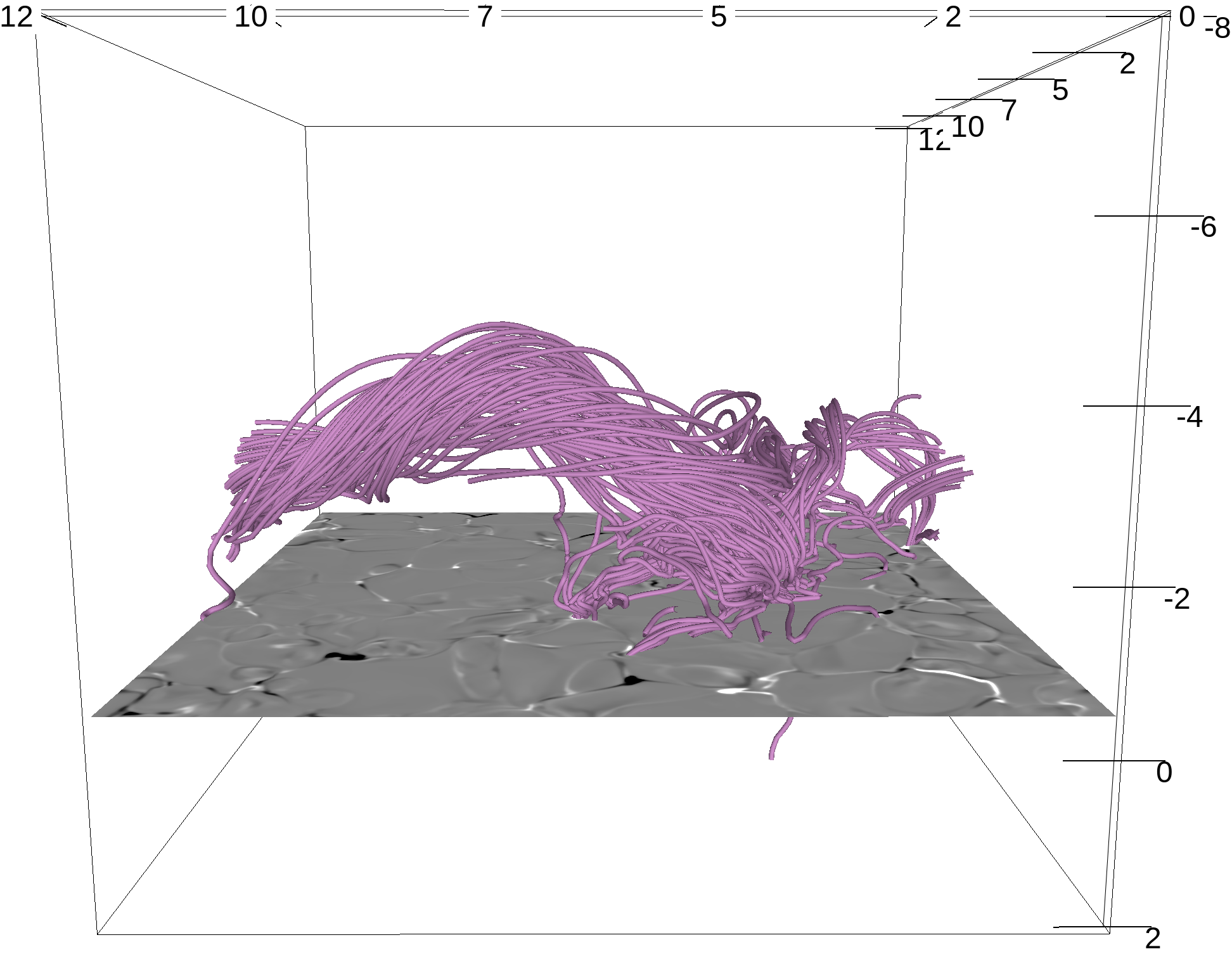}};
    \begin{scope}[x={(image.south east)},y={(image.north west)}]
    \node[black,fill=white,draw] at (0.3,0.9) {t* = 11\,040 s};
    \end{scope}
    \end{tikzpicture}
  % \subcaption[]{t* = 11\,040 s}
\end{subfigure}
\begin{subfigure}{.49 \textwidth}
  \centering
    \begin{tikzpicture}
    \node[anchor=south west,inner sep=0] (image) at (0,0)
        {\includegraphics[width=\linewidth]{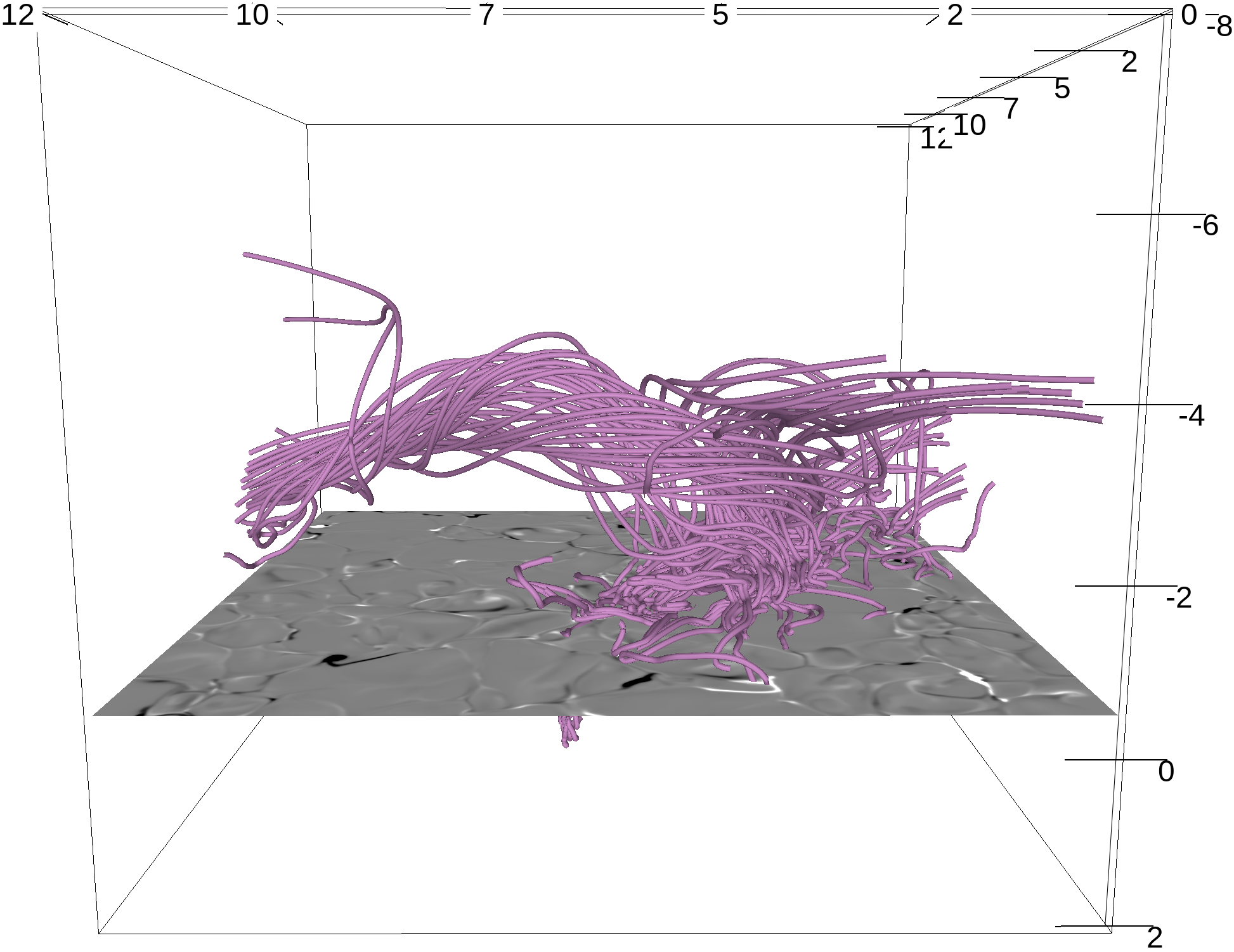}};
    \begin{scope}[x={(image.south east)},y={(image.north west)}]
    \node[black,fill=white,draw] at (0.3,0.9) {t* + 200 (t = 11\,240 s)};
    \end{scope}
    \end{tikzpicture}
  % \subcaption[]{t* + 200 (t = 11\,240 s)}
\end{subfigure}
\caption{Time series of the evolution of the flux rope at six different time stamps, with t* = 11\,040 s as the reference time.}
\label{fig:bigrope}
\end{figure*}
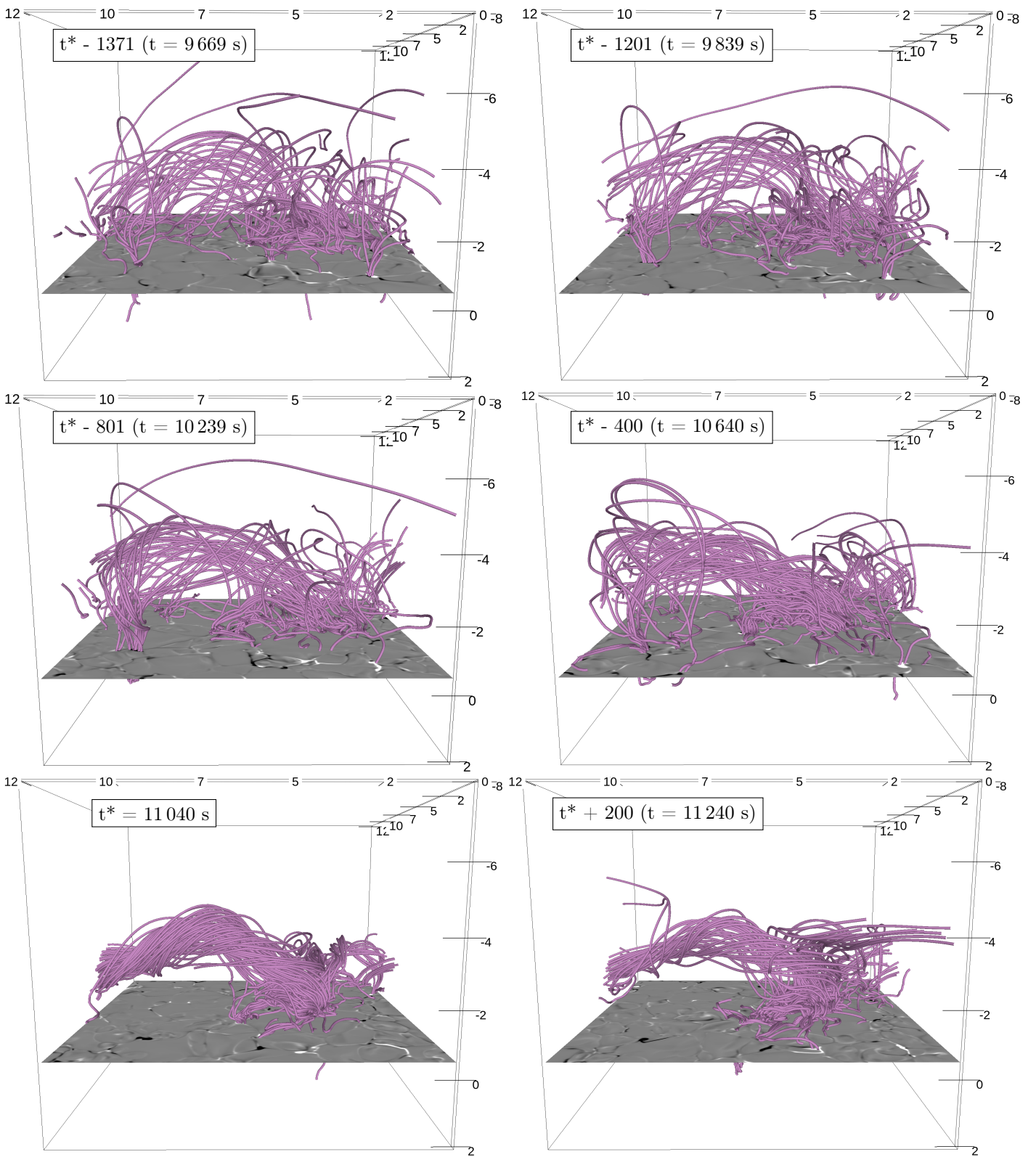

Equipped with the cork identifiers associated with each line in the flux rope, the history and future of the flux rope can be established because, assuming the field is frozen into the plasma except in diffusive regions, Lagrangian markers should trace consistent lines throughout the run. Under small-scale reconnection between line pairs, the cork acts as a seed point for whichever line passes through that coordinate; that is, a pair of corks that trace a pair of reconnecting lines will still trace the same pair of lines after reconnection. Figure \ref{fig:bigrope} illustrates a time series of the flux rope from corks seeded at t = 11\,040 s and moved forward and backward in time. The upper left panel is the flux rope at t = 9\,669 s, which is the first snapshot that we had Lagrangian markers in the simulation.

We note that in the upper and center panels of Figure \ref{fig:bigrope}, some of the lines have a photospheric root in a strong negative patch on the left side of the box. Once the lines on the left side of the flux rope have connected over the left horizontal boundary as in the lower panels of Figure \ref{fig:bigrope}, the rope becomes tightly ordered and eventually undergoes a major reconnection event as discussed in \citet{2022A&A...668A.177R}, but not shown here. 

From the evolution of the overall flux rope, it is clear that the formation of the rope is associated with thin current sheets inside and around it, and a major negative footpoint change from a strong root on the left side of the box to negative roots across the horizontal boundaries. This change in footpoint affinity points to systematic small-scale reconnection events within the flux rope itself, creating order out of the previously tangled lines. It follows that the formation of the overall flux rope depends on the relationship between reconnection in its coronal lines, and changes in their photospheric roots. In the following subsections, we use this overall flux system as a guide toward exploring even smaller scale ordering events, from smaller flux bundles within the wider flux rope down to component reconnection between individual magnetic field line pairs.

% \subsection{Changes in connectivity}
% As previously discussed, some of the lines that belong to the overall flux system begin initially 

\subsection{Merging flux bundles}\label{bundles}

During the onset of field ordering and subsequent large-scale reconnection, several quasi-separatrix layers (QSLs) exist within the flux rope \citep[see][]{1996A&A...308..643D}. This is evident by examining $\alpha$ values orthogonal to the flux rope as previously discussed, as well as determining where associated field lines are rooted in the photosphere and how that may change over time. In \citet{2022A&A...668A.177R}, we suggested that long-lasting photospheric roots are important for collecting and maintaining ordered fields, which is still the case. However, we have a network of convection-driven photospheric flux concentrations that all could be possible roots for the magnetic field. When we see coherent flux bundles diverge toward two different photospheric roots, it indicates the presence of a QSL further up the bundle and implies a reconnection event that had changed its structural coherence and photospheric connectivity. This is evidence of small-scale reconnection occurring within the flux rope; even though a bundle may be coherent as it passes through the reference slice, it is not necessarily coherent as it approaches the photosphere. It may meet another QSL on the way down, diverging to completely different roots. We see this behavior consistently throughout the formation of the flux rope, which suggests that small-scale reconnection must be occurring.

While the previous subsection discussed the shape and evolution of the overall flux rope, the number of QSLs and the ordering of the field from tangled flux bundles to coherent structure (as seen in Figure \ref{fig:bigrope}) suggests that the flux rope is self-ordering. Indeed, the overall twist of the flux system does not display consistent, large-scale helicity until just before the onset of major reconnection, and it builds up gradually from flux bundles beforehand. With that, it is possible that magnetic helicity cascades in this case not to smaller structures, but larger ones. We note that this flux rope is the result of spontaneous formation from coronal lines; we see no evidence that would support flux emergence, flux cancellation, or tether-cutting reconnection here.

\begin{figure*}
\begin{subfigure}{.49\textwidth}
  \centering
    \begin{tikzpicture}
    \node[anchor=south west,inner sep=0] (image) at (0,0)
        {\includegraphics[width=0.95\linewidth]{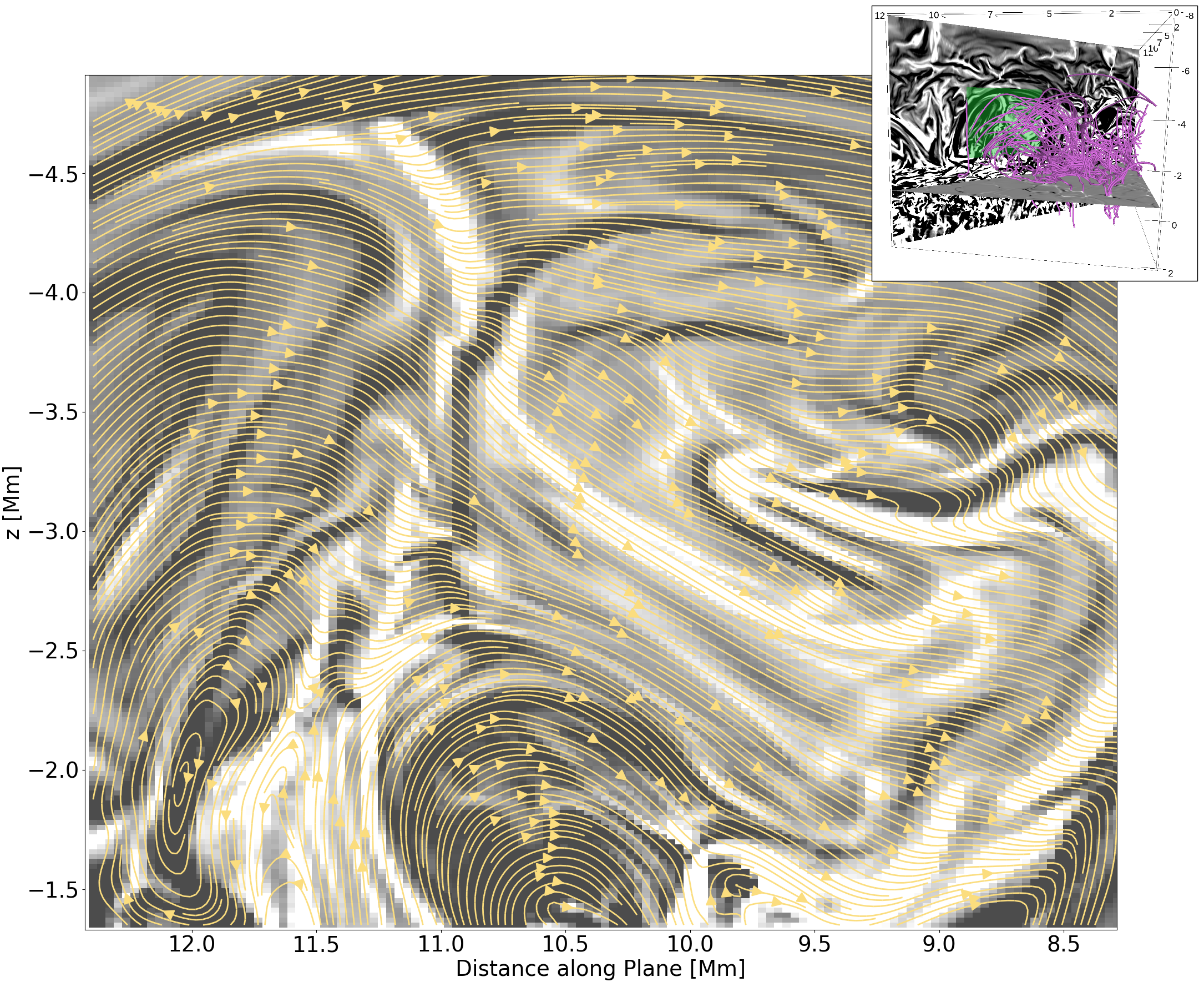}};
    \begin{scope}[x={(image.south east)},y={(image.north west)}]
    \node[black,fill=white,draw] at (0.2,0.95) {t = 9\,739 s};
    \end{scope}
    \end{tikzpicture}
  % \subcaption[]{t = 11\,040 s}
  \label{fig:con1130}
\end{subfigure}
\begin{subfigure}{.49\textwidth}
  \centering
    \begin{tikzpicture}
    \node[anchor=south west,inner sep=0] (image) at (0,0)
        {\includegraphics[width=0.95\linewidth]{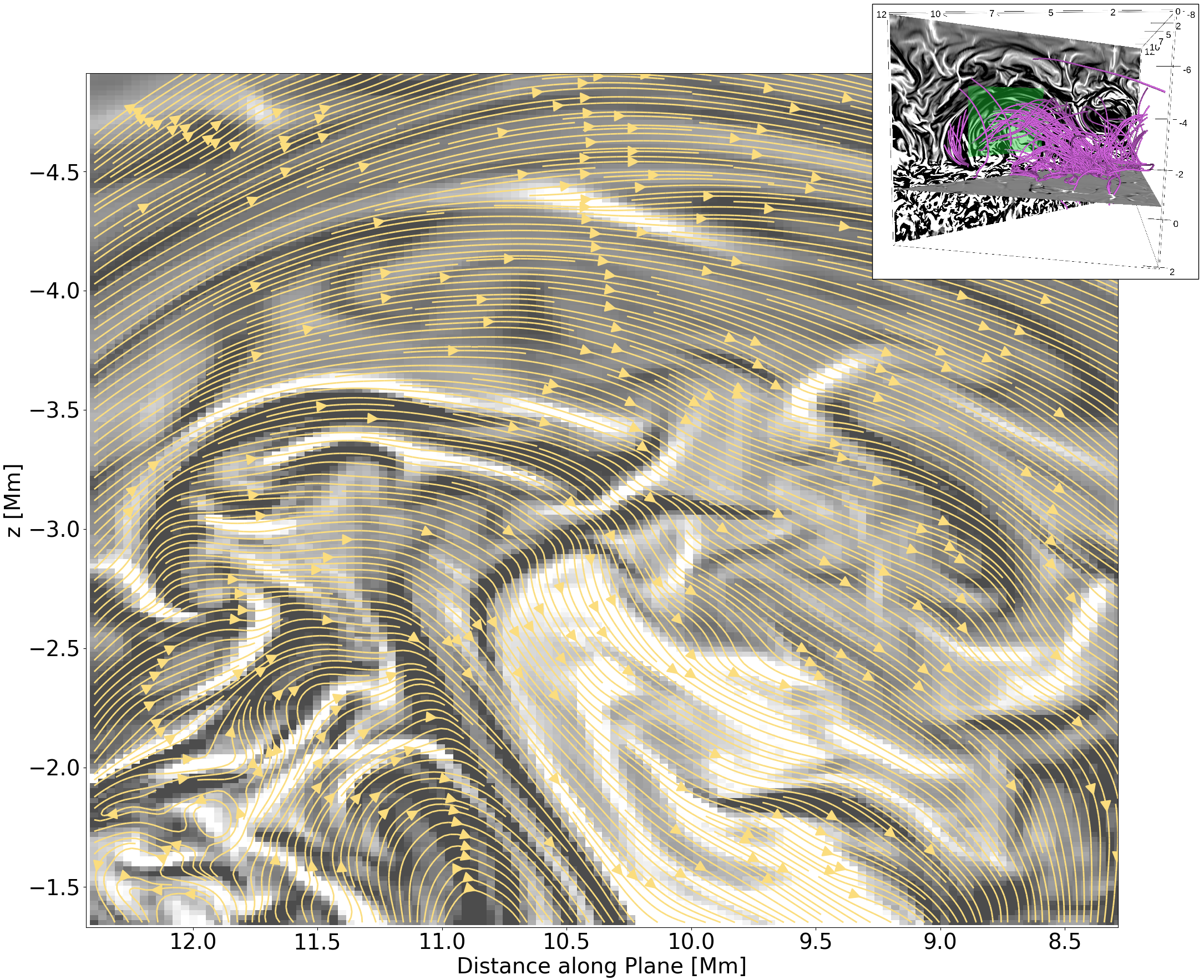}};
    \begin{scope}[x={(image.south east)},y={(image.north west)}]
    \node[black,fill=white,draw] at (0.2,0.95) {t = 10\,239 s};
    \end{scope}
    \end{tikzpicture} 
  % \subcaption[]{t = 11\,080 s}
  \label{fig:con1200}
  \end{subfigure}
\begin{subfigure}{.49\textwidth}
  \centering
    \begin{tikzpicture}
    \node[anchor=south west,inner sep=0] (image) at (0,0)
        {\includegraphics[width=0.95\linewidth]{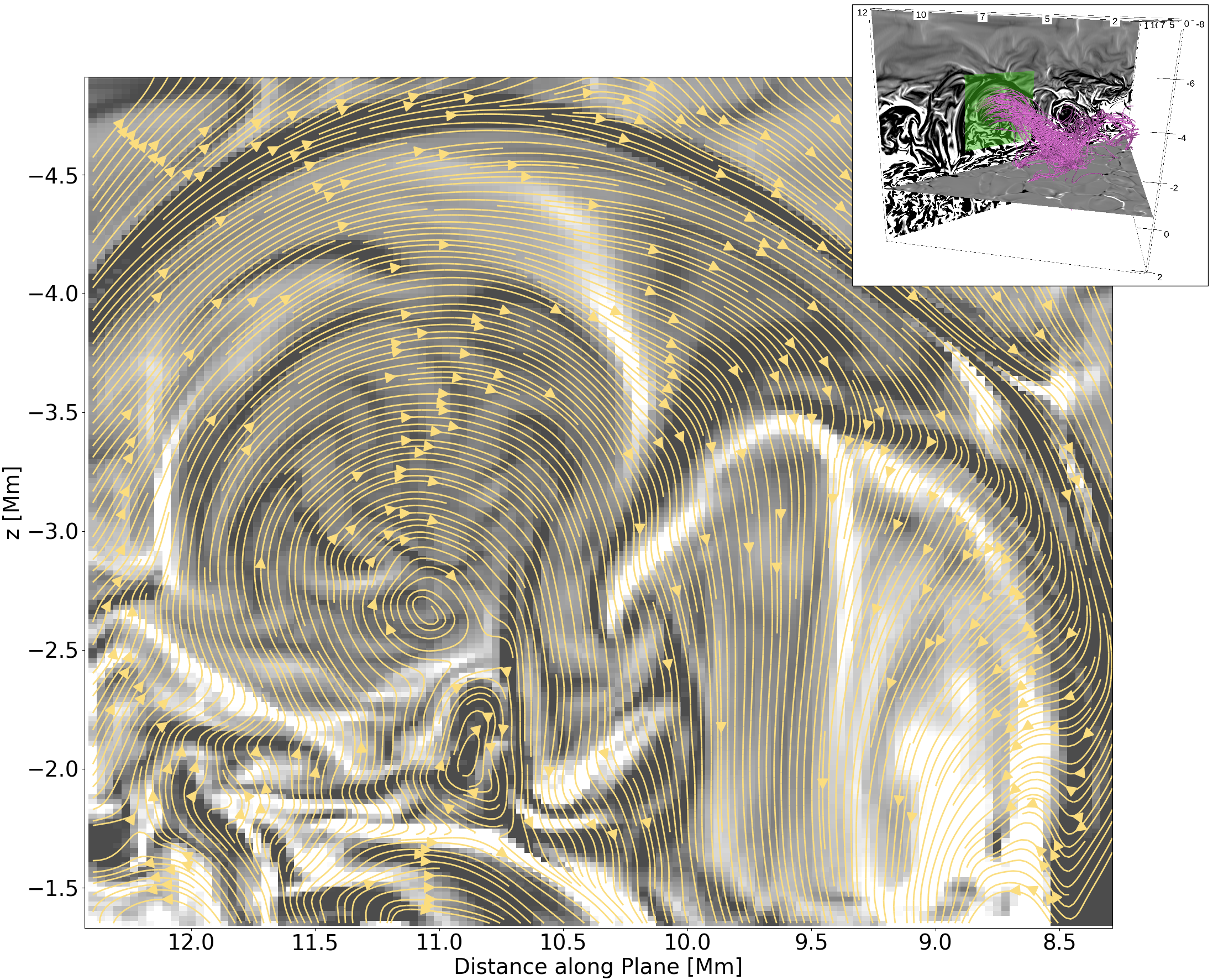}};
    \begin{scope}[x={(image.south east)},y={(image.north west)}]
    \node[black,fill=white,draw] at (0.2,0.95) {t = 11\,040 s};
    \end{scope}
    \end{tikzpicture}
  % \subcaption[]{t = 11\,040 s}
  \label{fig:con1260}
\end{subfigure}
\begin{subfigure}{.49\textwidth}
  \centering
    \begin{tikzpicture}
    \node[anchor=south west,inner sep=0] (image) at (0,0)
        {\includegraphics[width=0.95\linewidth]{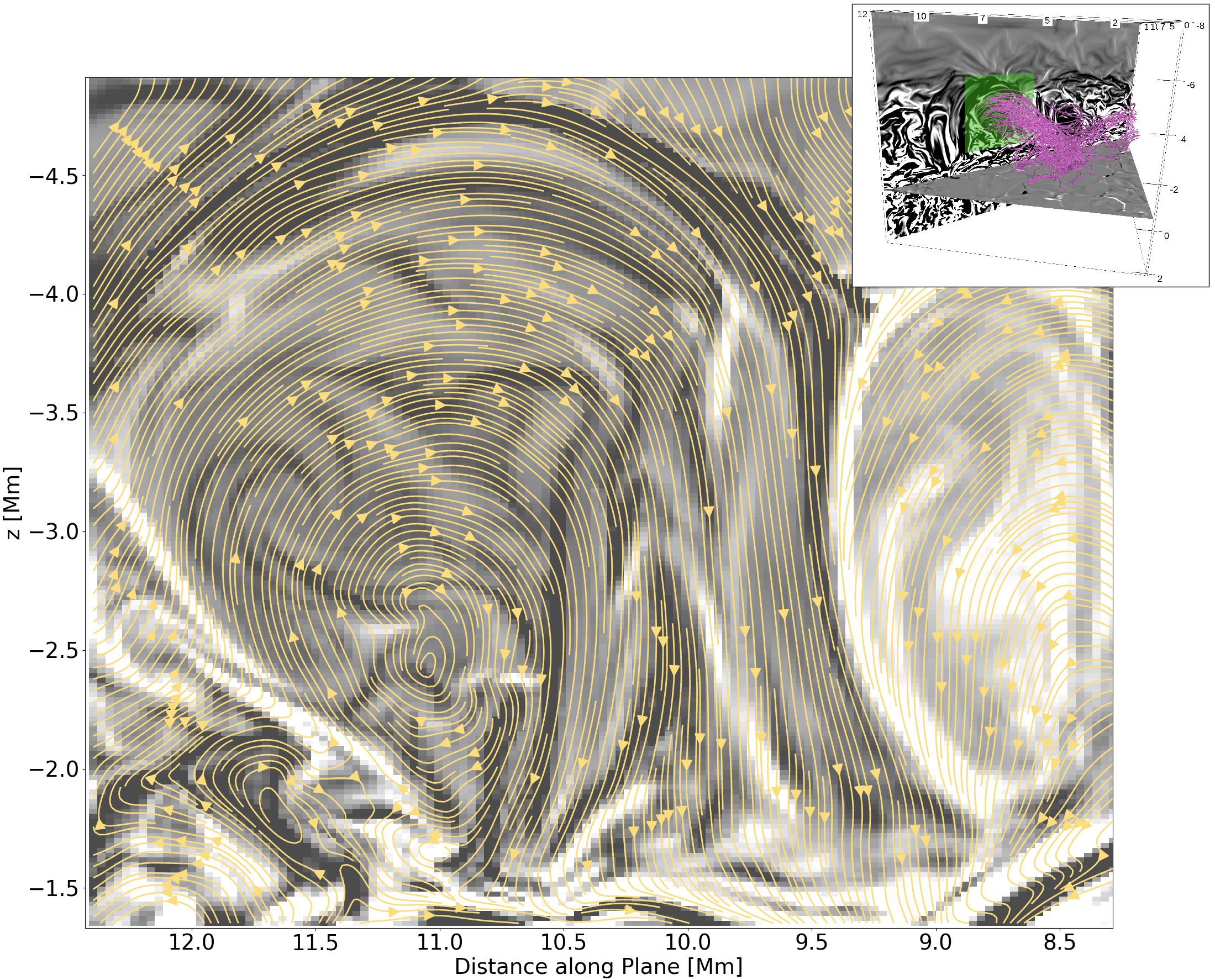}};
    \begin{scope}[x={(image.south east)},y={(image.north west)}]
    \node[black,fill=white,draw] at (0.2,0.95) {t = 11\,080 s};
    \end{scope}
    \end{tikzpicture} 
  % \subcaption[]{t = 11\,080 s}
  \label{fig:con1264}
\end{subfigure}
\begin{subfigure}{.49\textwidth}
  \centering
    \begin{tikzpicture}
    \node[anchor=south west,inner sep=0] (image) at (0,0)
        {\includegraphics[width=0.95\linewidth]{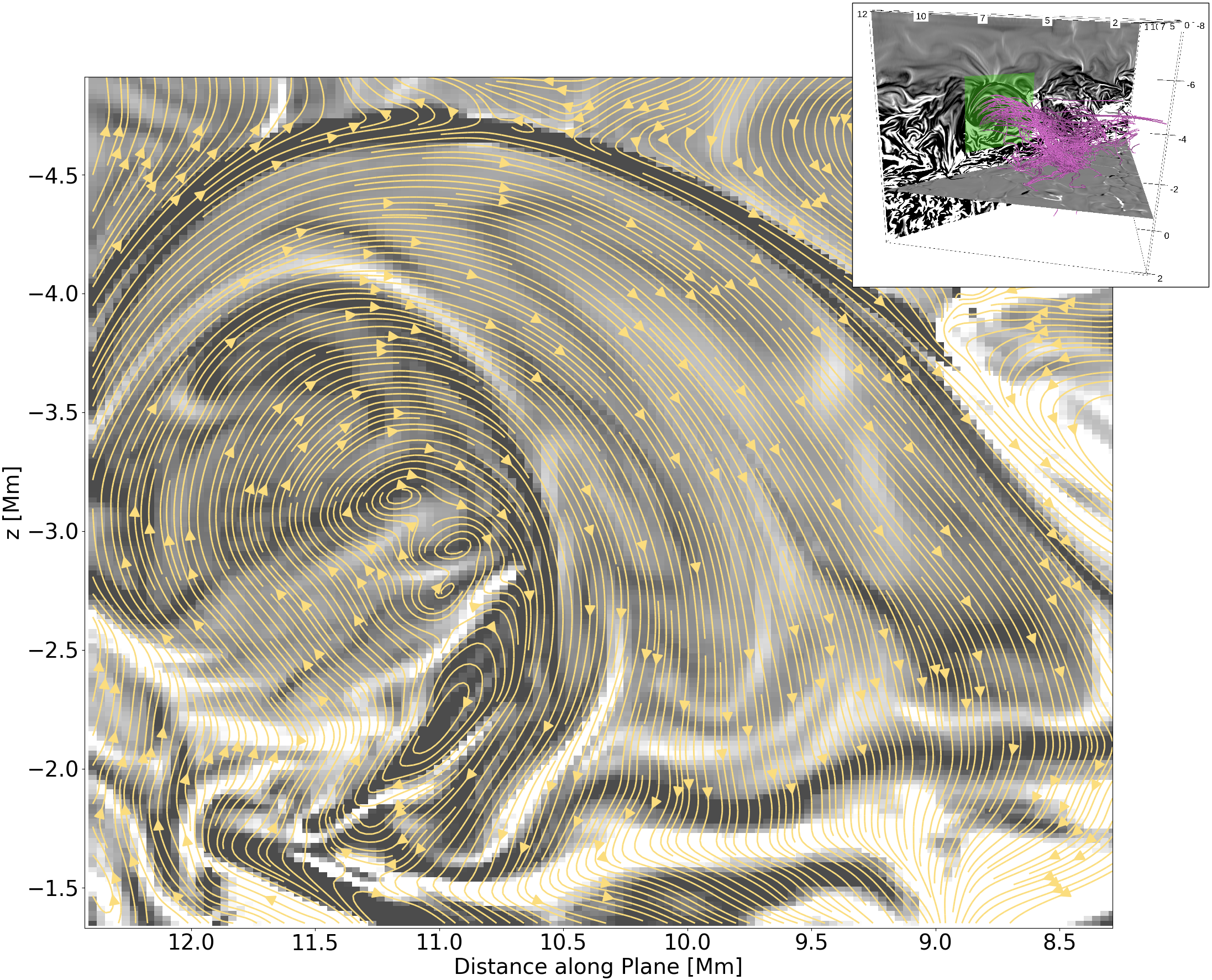}};
    \begin{scope}[x={(image.south east)},y={(image.north west)}]
    \node[black,fill=white,draw] at (0.2,0.95) {t = 11\,190 s};
    \end{scope}
    \end{tikzpicture} 
  % \subcaption[]{t = 11\,190 s}
  \label{fig:con1275}
\end{subfigure}
\begin{subfigure}{.49\textwidth}
  \centering
    \begin{tikzpicture}
    \node[anchor=south west,inner sep=0] (image) at (0,0)
        {\includegraphics[width=0.95\linewidth]{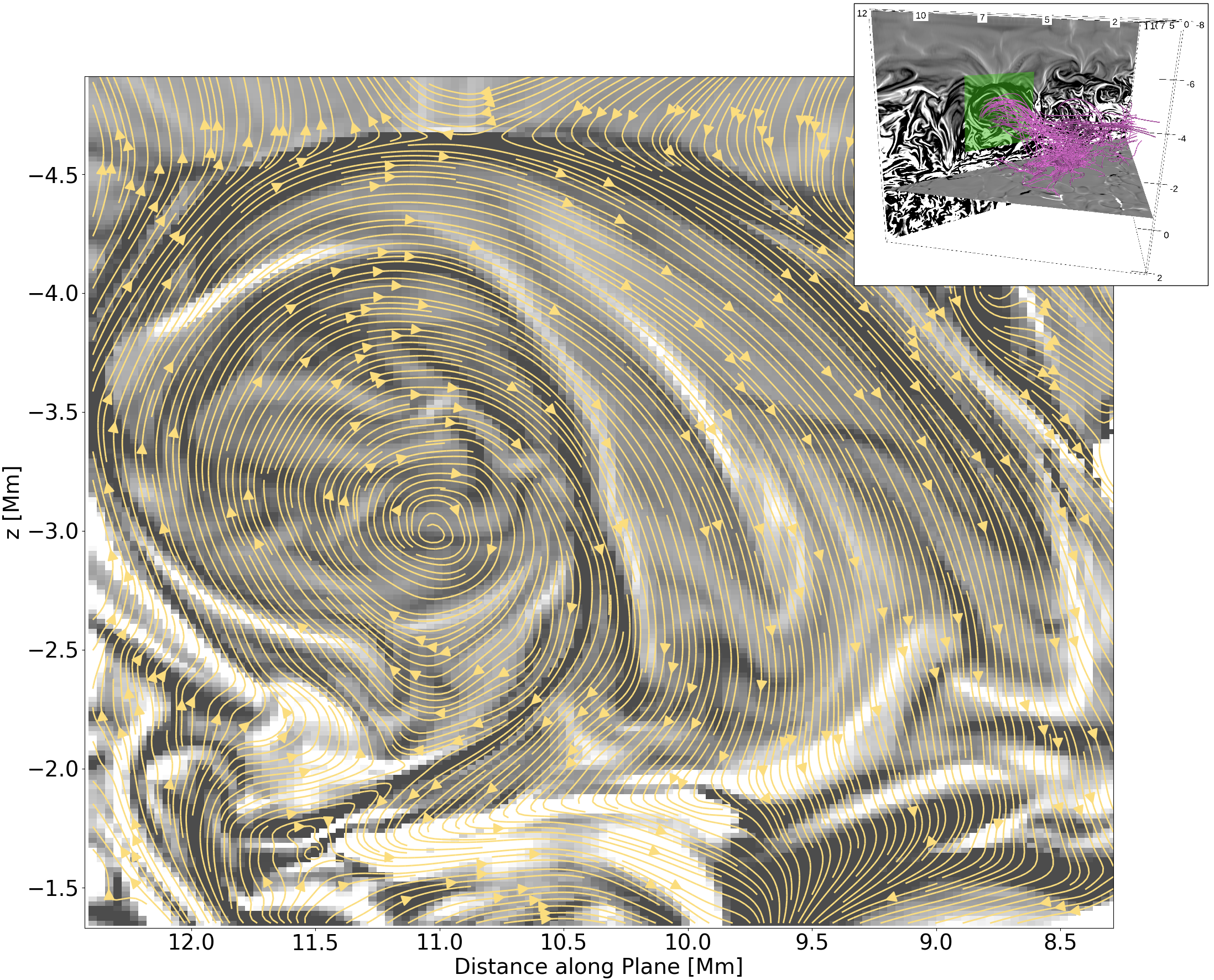}};
    \begin{scope}[x={(image.south east)},y={(image.north west)}]
    \node[black,fill=white,draw] at (0.2,0.95) {t = 11\,240 s};
    \end{scope}
    \end{tikzpicture}  
  % \subcaption[]{t = 11\,240 s}
  \label{fig:con1280}
\end{subfigure}
\begin{subfigure}{.99\textwidth}
  \centering
  \includegraphics[width=0.5\linewidth]{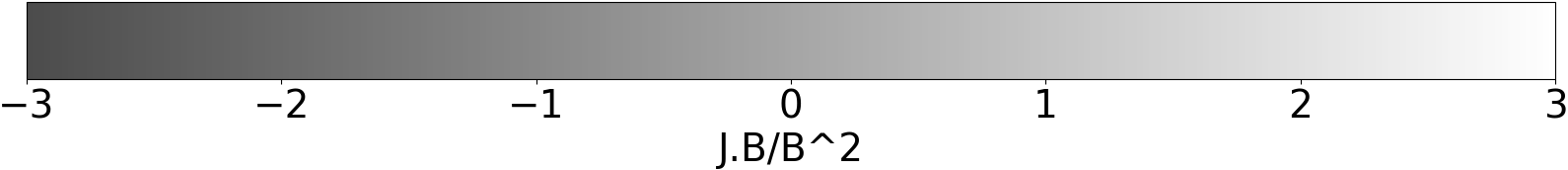}
  % \subcaption[]{t = 11\,040 s}
\end{subfigure}
\caption{Vertical slices showing $\alpha$ (Equation~\ref{eq:1}) along an orthogonal plane with respect to the flux rope. Yellow contours represent the field strength along the plane with arrows indicating the direction of twist. Each panel represents a different timestep, and each includes a 3D inset showing the vertical plane colored by $\alpha$ and showing flux rope lines. The green box on the inset represents the size and location of the contour plot.}
\label{fig:contours}
\end{figure*}

To demonstrate the inverse cascade of magnetic helicity, we must find small-scale twisting ropes which merge into a larger twisting rope, separated by a QSL. These can be isolated by inspecting a 2D slice through the flux rope and searching for contours in the magnetic field. Figure \ref{fig:contours} shows a vertical slice through the flux rope (the same vertical slice as in Figure \ref{fig:flux_compare}, but cropped to the scale of the rope) with superimposed magnetic field contours shown in yellow. For reference, see the small 3D inset located at the upper right of each panel. 

The upper two panels of Figure \ref{fig:contours} show the same region before the flux rope forms, in order to see that no relevant flux system exists at those early time stamps. The center panels of Figure \ref{fig:contours} show a before (left) and after (right) merger between two small-scale twisting ropes into one larger one. These two ropes, evidenced by small contours, twist in the same direction and are separated by a small current sheet. As demonstrated, the two flux bundles merge between t = 11\,040 s and t = 11\,080 s. 

Although the center right panel represents the merger of those two flux bundles, it does not represent the final helical state. The lower two panels of Figure \ref{fig:contours} show another similar event; the lower left panel showing four distinct bundles at t = 11\,190 s and the lower right panel showing the final helicity at t = 11\,240 s before the major reconnection event. The four bundles are, again, initially separated by a system of current sheets that become far less pronounced after the merger. 

\begin{figure*}
\begin{subfigure}{.5\textwidth}
  \centering
  % include first image
      \begin{tikzpicture}
    \node[anchor=south west,inner sep=0] (image) at (0,0)
        {\includegraphics[width=\linewidth]{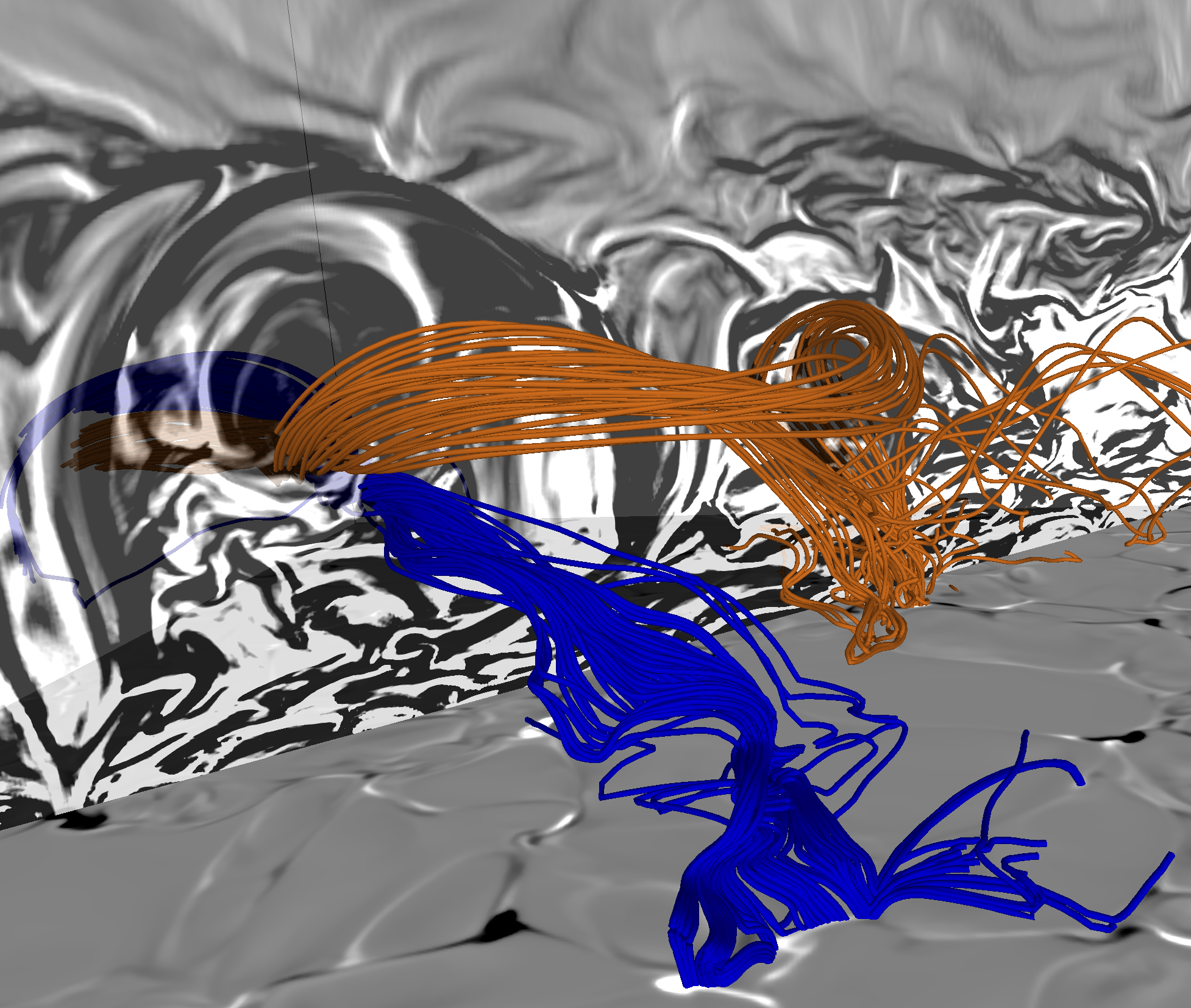}};
    \begin{scope}[x={(image.south east)},y={(image.north west)}]
    \node[black,fill=white,draw] at (0.2,0.95) {t = 11\,040 s};
    \end{scope}
    \end{tikzpicture} 
  % \includegraphics[width=\linewidth]{FIGURES/closeup_two_twists_1260.png}
  % \subcaption[]{t = 11\,040 s}
\end{subfigure}
\begin{subfigure}{.5\textwidth}
  \centering
  % include second image
    \begin{tikzpicture}
\node[anchor=south west,inner sep=0] (image) at (0,0)
    {\includegraphics[width=\linewidth]{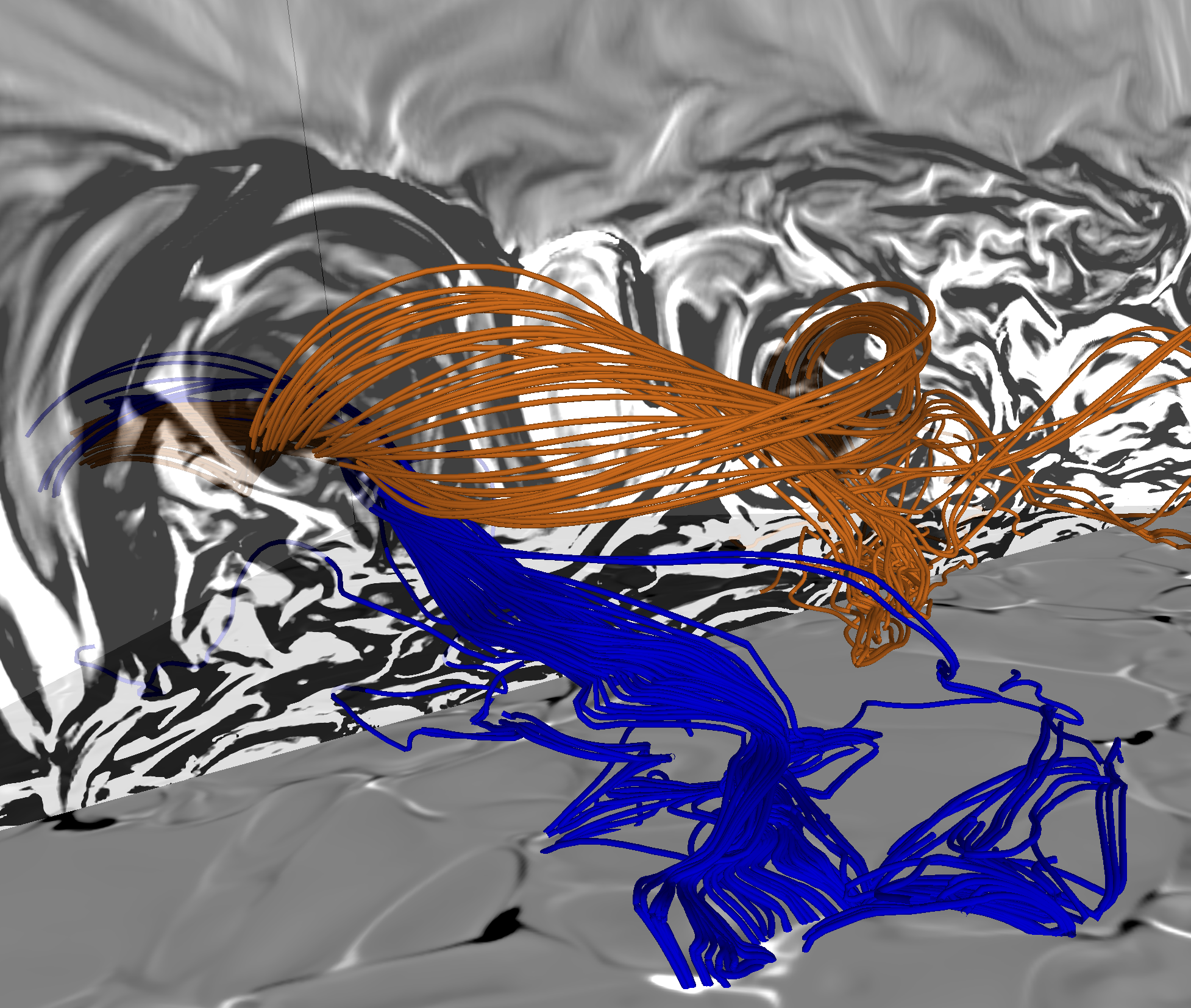}};
\begin{scope}[x={(image.south east)},y={(image.north west)}]
\node[black,fill=white,draw] at (0.2,0.95) {t = 11\,080 s};
\end{scope}
\end{tikzpicture} 
  % \includegraphics[width=\linewidth]{FIGURES/closeup_two_twists_1264.png}  
  % \subcaption[]{t = 11\,080 s}
\end{subfigure}
\begin{subfigure}{.5\textwidth}
  \centering
  % include first image
        \begin{tikzpicture}
    \node[anchor=south west,inner sep=0] (image) at (0,0)
        {\includegraphics[width=\linewidth]{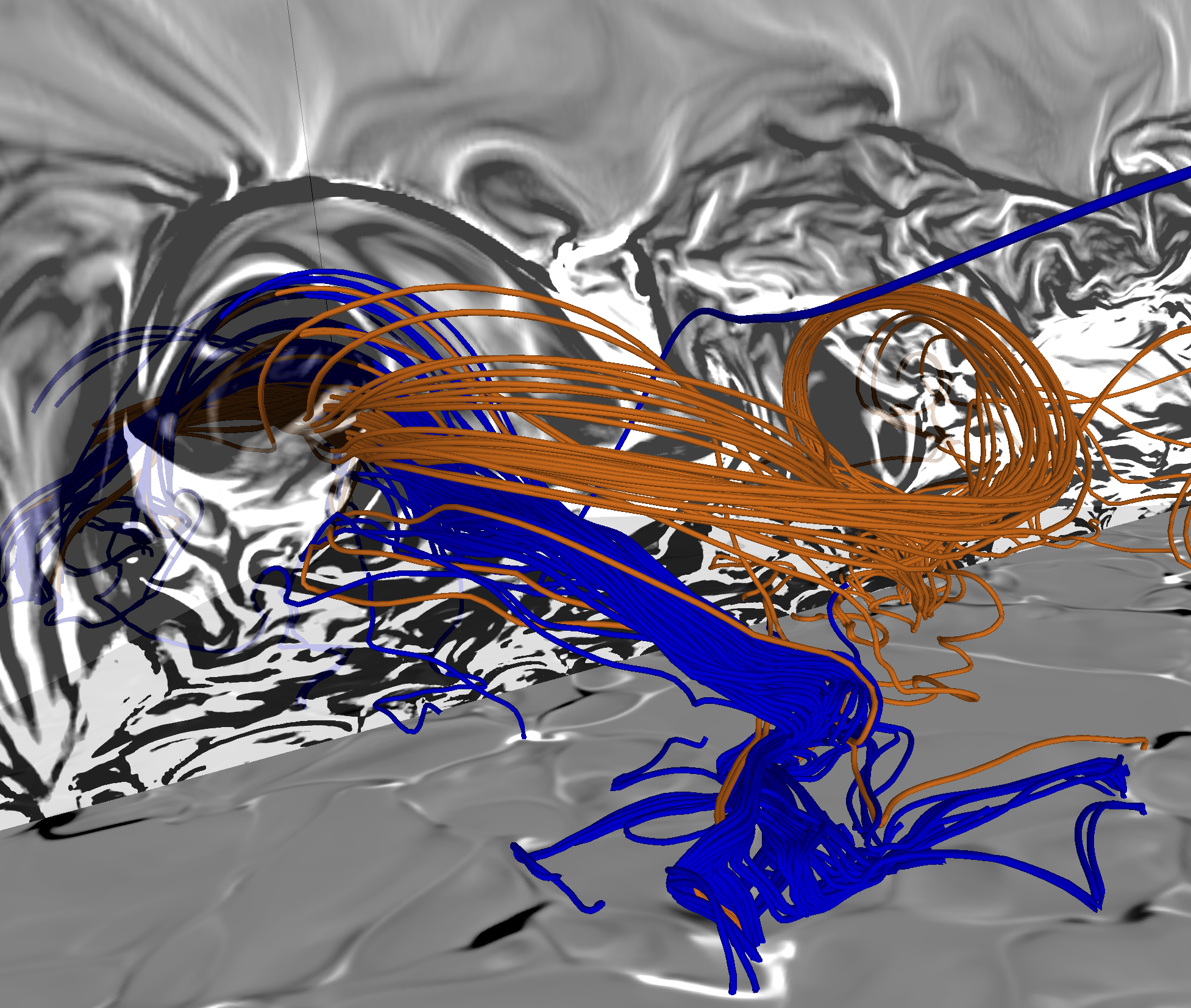}};
    \begin{scope}[x={(image.south east)},y={(image.north west)}]
    \node[black,fill=white,draw] at (0.2,0.95) {t = 11\,190 s};
    \end{scope}
    \end{tikzpicture} 
  % \includegraphics[width=\linewidth]{FIGURES/closeup_two_twists_1275.png}
  % \subcaption[]{t = 11\,190 s}
\end{subfigure}
\begin{subfigure}{.5\textwidth}
  \centering
  % include second image
      \begin{tikzpicture}
    \node[anchor=south west,inner sep=0] (image) at (0,0)
        {\includegraphics[width=\linewidth]{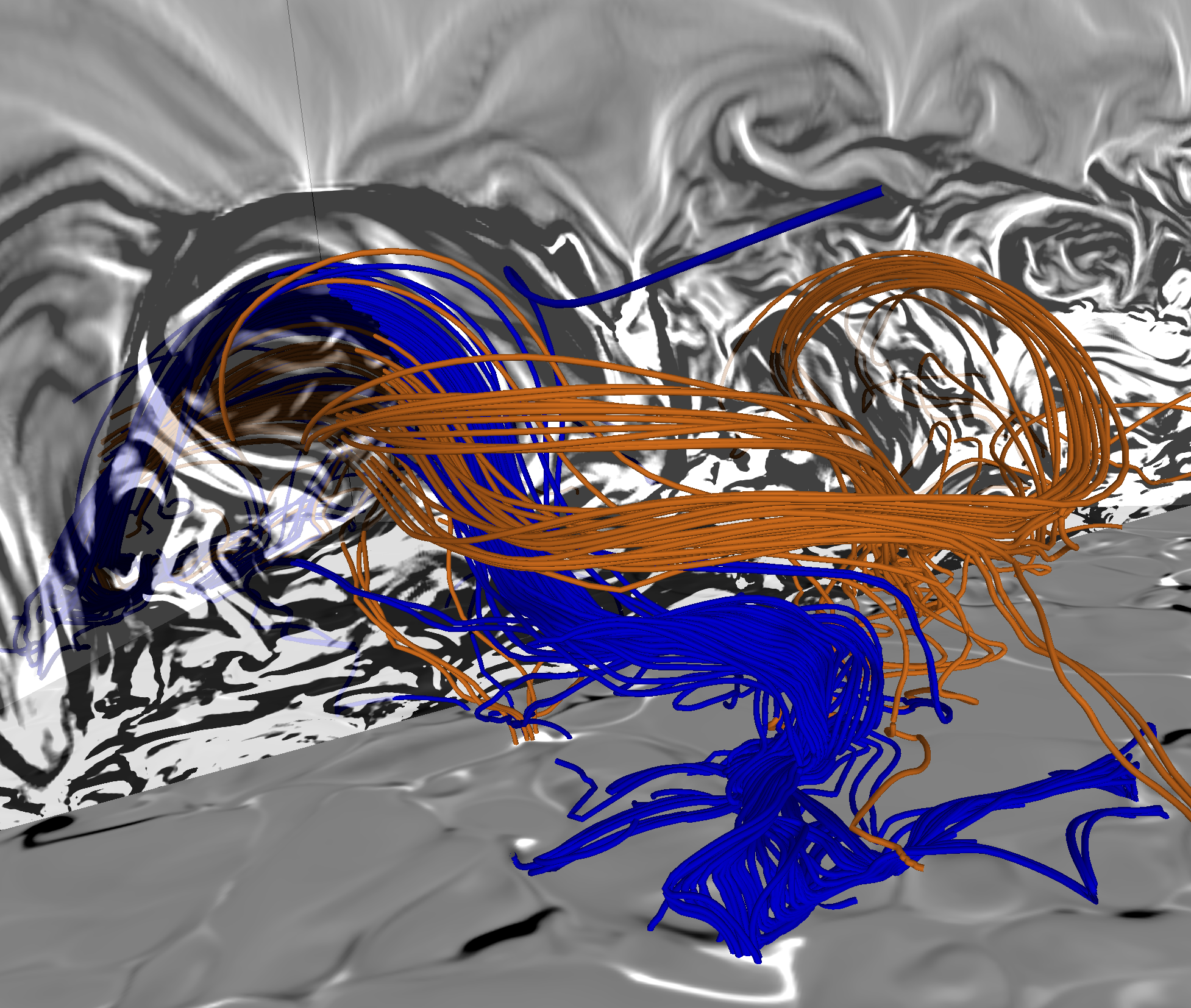}};
    \begin{scope}[x={(image.south east)},y={(image.north west)}]
    \node[black,fill=white,draw] at (0.2,0.95) {t = 11\,240 s};
    \end{scope}
    \end{tikzpicture} 
  % \includegraphics[width=\linewidth]{FIGURES/closeup_two_twists_1280.png}  
  % \subcaption[]{t = 11\,240 s}
\end{subfigure}
\caption{Zoomed-in 3D renderings of the flux bundles associated with the contours shown in the center and lower panels of Figure \ref{fig:contours}.}
\label{fig:closeup}
\end{figure*}

For a closer look at the mergers between flux bundles, Figure \ref{fig:closeup} illustrates the evolution of the two flux bundles associated with the two small contours shown in the center left panel of Figure \ref{fig:contours}. Figure \ref{fig:closeup} shows a close-up 3D rendering of the two flux bundles (orange and blue) with respect to where they bisect the vertical cut at the same time stamps as in the center and lower panels of Figure \ref{fig:contours}: t = 11\,040 s (center left), t = 11\,080 s (center right), t = 11\,190 s (lower left), and t = 11\,240 s (lower right). The field lines were seeded at t = 11\,040 s by a random distribution of seeds within two small cubes located on either side of the major current sheet separating the contours. Short field line segments were bidirectionally drawn from those seeds, and then the corks spatially nearest to each line segment were selected and used to seed the full lines at each time stamp. As noted in Section \ref{cx}, the corks are necessary to provide a reliable time series such that the integrator is always tracing lines seeded by the same set of Lagrangian markers in time.

The orange and blue flux bundles illustrated in Figure \ref{fig:closeup} begin as two distinct and fully separate bundles, but evolve over time into a wider, more relaxed part of the overall flux system. As evidenced by changes in photospheric connectivity as well as the known existence of thin current sheets, the blue and orange bundles evolve via small-scale reconnection and end up tangled around one another within the flux rope. We note that while it is possible for flux bundles to undergo tilt or kink-tilt instabilities \citep{1990PhFlB...2..488R,2014ApJ...795...77K}, we see here that our flux bundles are too weakly twisted to observe this.

It is clear from Figure \ref{fig:closeup} that the bundles initially cross the vertical slice toward the bottom of the flux rope and eventually twist their way around as they undergo small-scale reconnection, merge, and evolve. We note that at t = 11\,190 s, the lines from the original two bundles also pass through the four new contours (lower left panel of Figure \ref{fig:contours}) and then continue to evolve to the helical state at t = 11\,240 s. This means that the same lines that comprise the two bundles at t = 11\,040 s are also some of the lines that comprise the four bundles at t = 11\,190 s. The lower left panel of Figure \ref{fig:closeup} shows lines passing through the vertical slice in between the same current sheets that separate the four contours. This means that the blue and orange lines merge, split into four bundles, then finally merge again to the helical state at t = 11\,240 s, just before the onset of the major reconnection event.

By following the orange and blue lines, we see coalescence of separate flux bundles into a final coherent flux system via a combination of relaxation and small-scale reconnection. This coalescence of bundles via small-scale reconnection is a result of convective driving and, potentially, the coalescence instability \citep{zhukov,2018PhPl...25h2904M}. We also see the removal of small-scale structures in favor of, eventually, a large-scale helical system. This illustrates the inverse cascade of helicity via small-scale reconnection, but over a set of two bulk flux systems within the wider flux rope. Next, we must consider the sources of small-scale reconnection by illustrating examples of individual magnetic field line pairs that undergo component reconnection. 

\subsection{Component reconnection along field line pairs}\label{component}
As we discussed in the previous subsection, we see evidence of reconnection in the bulk evolution of two distinct merging flux bundles. To support the idea that these flux systems are indeed reconnecting within one another, we present evidence of component reconnection along several field line pairs that are members of the aforementioned flux systems and cross the vertical cut in or around thin current sheets that appear at t = 11\,190 s. Here, component reconnection refers to reconnection dominated by the guide field, between a pair of lines that are almost parallel but slightly offset from one another \citep{1976JGR....81.3455C,2002JGRA..107.1332M,2007JGRA..112.8210T}. A pair of lines undergoing component reconnection will effectively switch places, so searching for such pairs requires searching for lines that suddenly switch footpoints. 

\begin{figure*}
\begin{subfigure}{.49\textwidth}
  \centering
  % include first image
      \begin{tikzpicture}
    \node[anchor=south west,inner sep=0] (image) at (0,0)
        {\includegraphics[width=\linewidth]{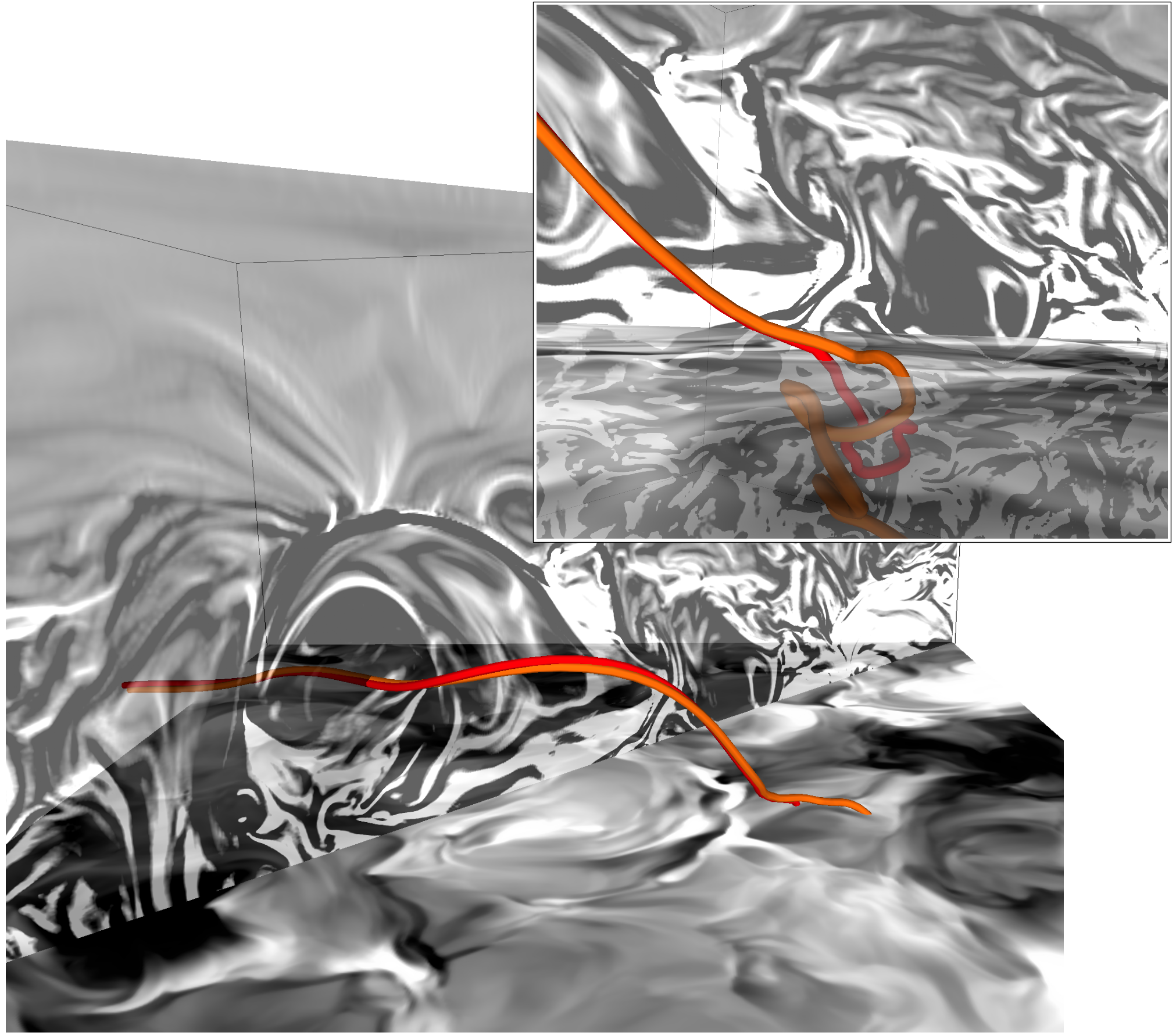}};
    \begin{scope}[x={(image.south east)},y={(image.north west)}]
    \node[black,fill=white,draw] at (0.2,0.9) {t = 11\,190 s};
    \end{scope}
    \end{tikzpicture} 
  % \includegraphics[width=\linewidth]{FIGURES/sel2_inset_1275.png}
  % \subcaption[]{t = 11\,190 s}
\end{subfigure}
\begin{subfigure}{.49\textwidth}
  \centering
  % include first image
      \begin{tikzpicture}
    \node[anchor=south west,inner sep=0] (image) at (0,0)
        {\includegraphics[width=\linewidth]{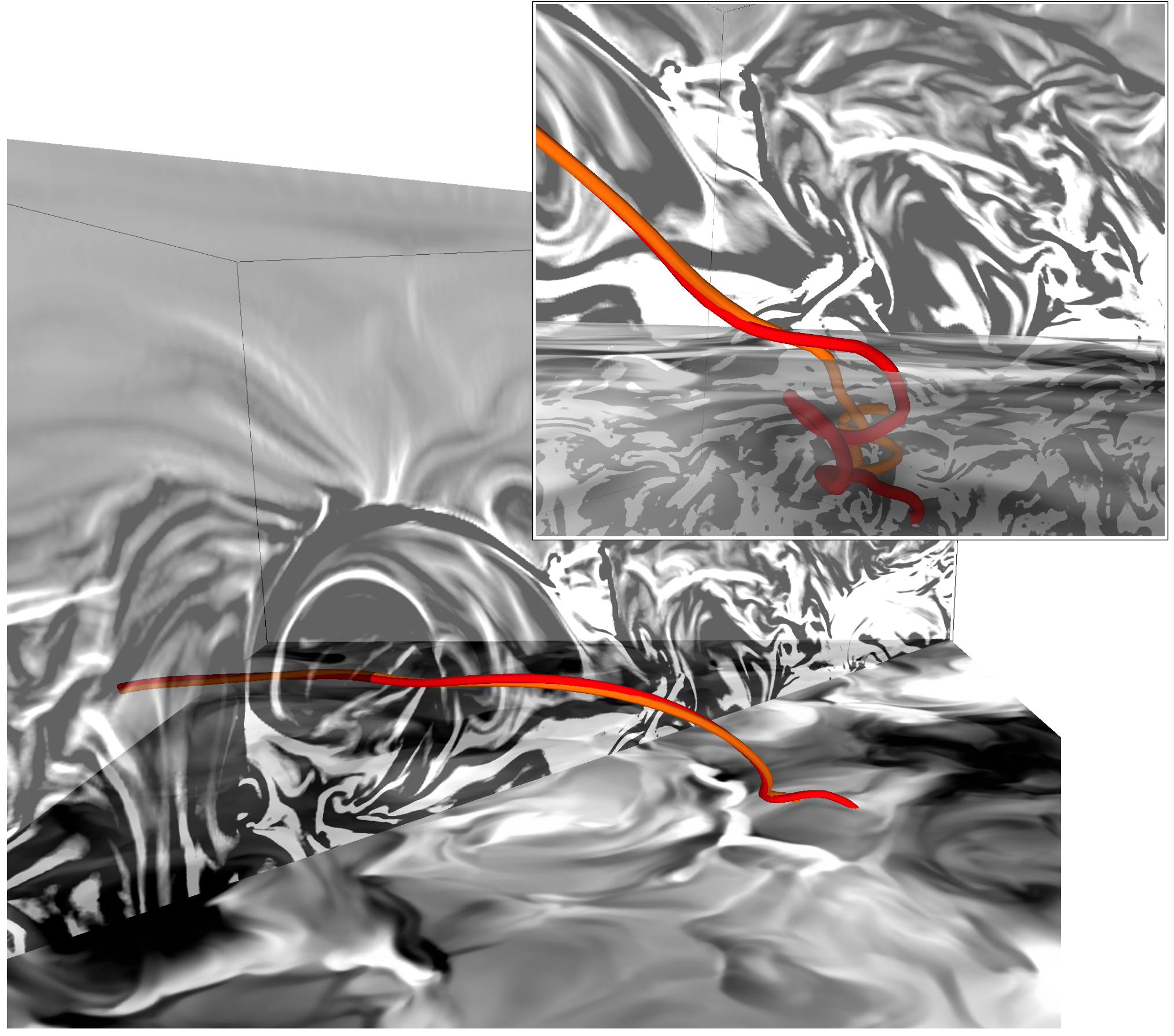}};
    \begin{scope}[x={(image.south east)},y={(image.north west)}]
    \node[black,fill=white,draw] at (0.2,0.9) {t = 11\,200 s};
    \end{scope}
    \end{tikzpicture} 
  % \includegraphics[width=\linewidth]{FIGURES/sel2_inset_1276.png}
  % \subcaption[]{t = 11\,200 s}
\end{subfigure}
    \caption{Field line pair at t = 11\,190 s (left) and t = 11\,200 s (right) that are traced by the same two Lagrangian markers, but switch chromospheric footpoints between the two time stamps. The insets in the upper right corners offer a zoomed-in view of the footpoint switching at a different viewing angle, as evidenced by the lines switching colors.}
    \label{fig:sel2}
\end{figure*}

\begin{figure*}
\begin{subfigure}{.5\textwidth}
  \centering
  % include first image
      \begin{tikzpicture}
    \node[anchor=south west,inner sep=0] (image) at (0,0)
        {\includegraphics[width=\linewidth]{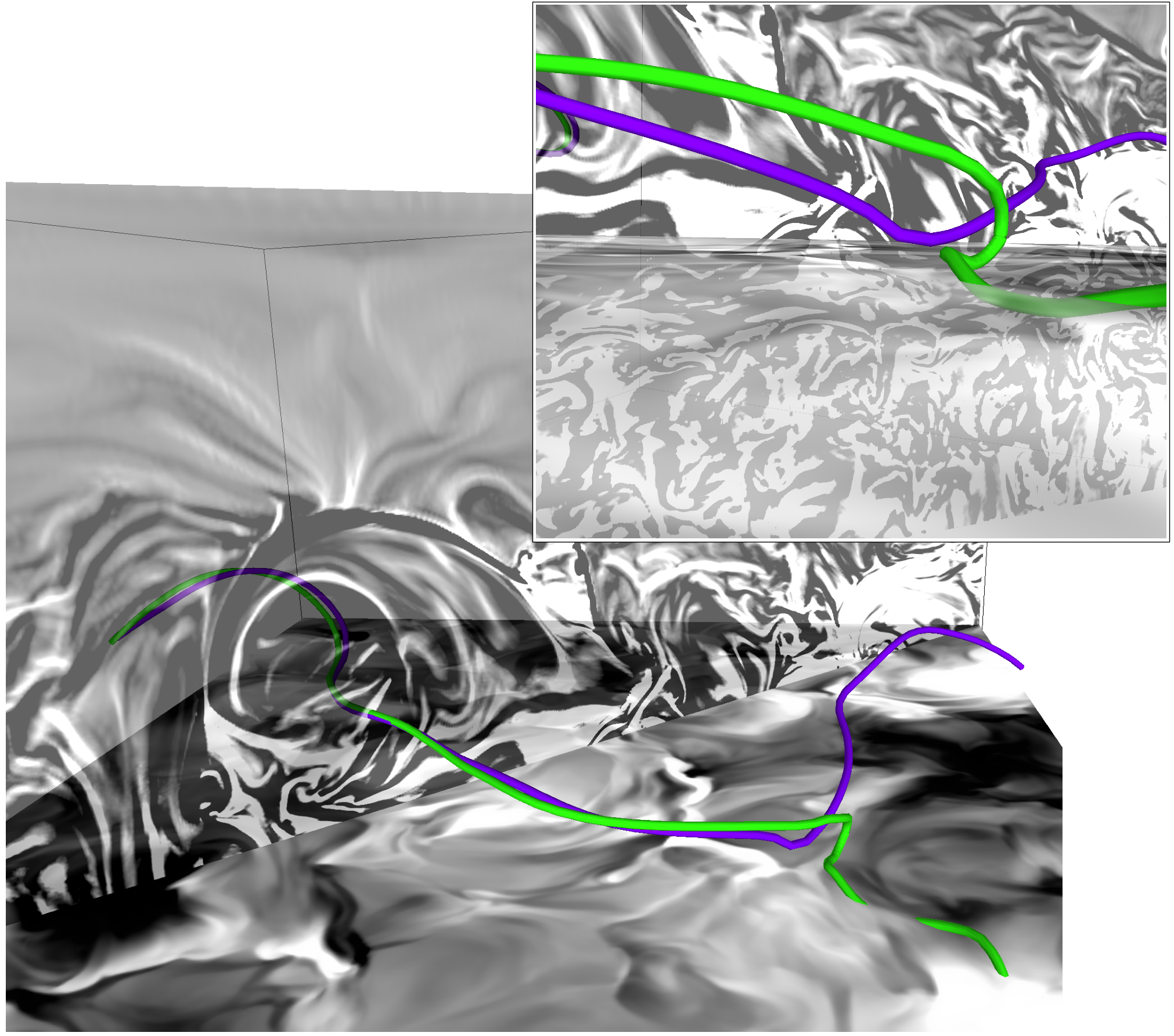}};
    \begin{scope}[x={(image.south east)},y={(image.north west)}]
    \node[black,fill=white,draw] at (0.2,0.9) {t = 11\,200 s};
    \end{scope}
    \end{tikzpicture} 
  % \includegraphics[width=\linewidth]{FIGURES/sel5_inset_1276.png}
  % \subcaption[]{t = 11\,200 s}
\end{subfigure}
\begin{subfigure}{.5\textwidth}
  \centering
  % include first image
      \begin{tikzpicture}
    \node[anchor=south west,inner sep=0] (image) at (0,0)
        {\includegraphics[width=\linewidth]{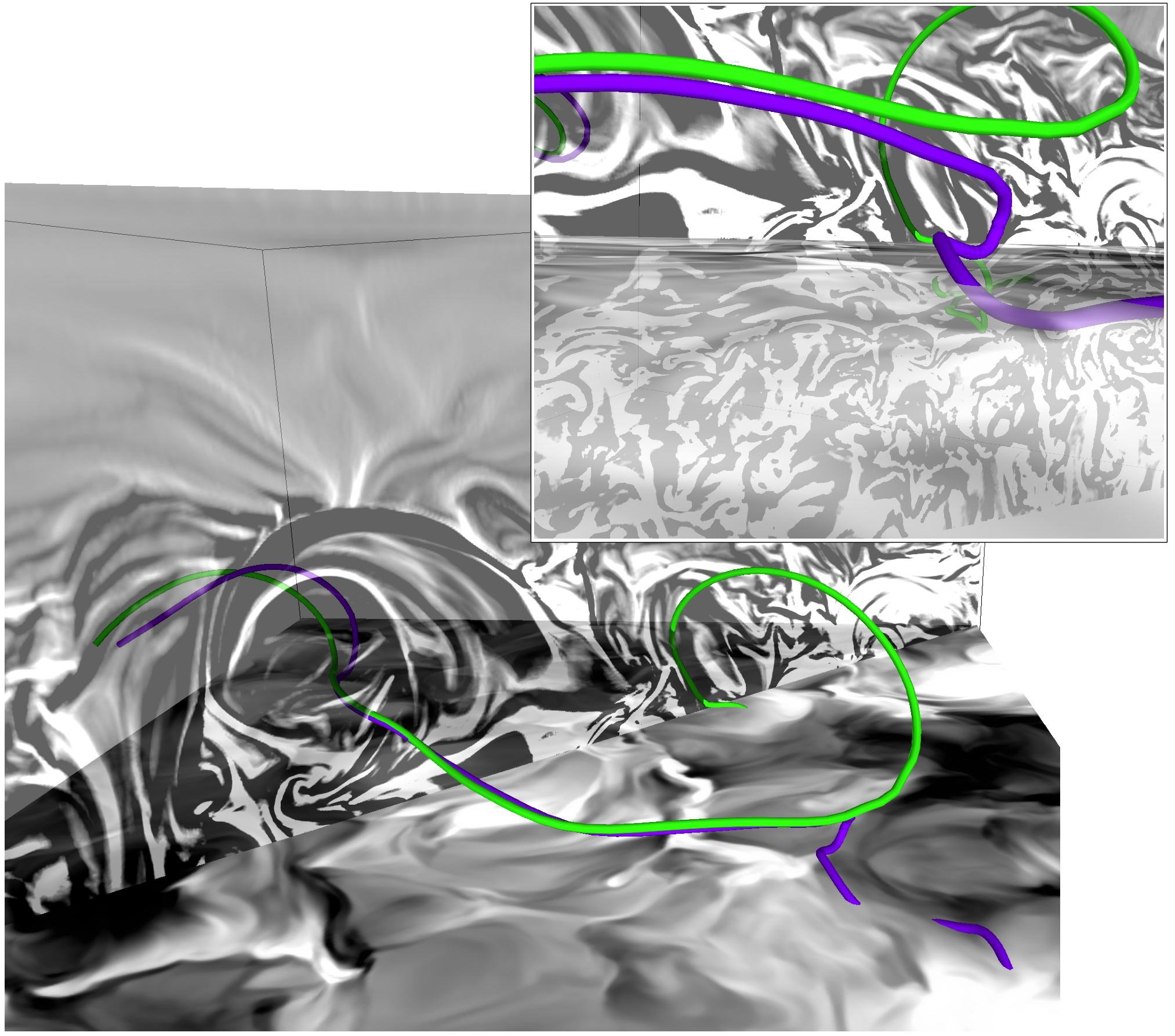}};
    \begin{scope}[x={(image.south east)},y={(image.north west)}]
    \node[black,fill=white,draw] at (0.2,0.9) {t = 11\,210 s};
    \end{scope}
    \end{tikzpicture} 
  % \includegraphics[width=\linewidth]{FIGURES/sel5_inset_1277.png}
  % \subcaption[]{t = 11\,210 s}
\end{subfigure}
    \caption{Field line pair at t = 11\,200 s (left) and t = 11\,210 s (right) that are traced by the same two Lagrangian markers, but switch chromospheric footpoints between the two time stamps. The insets in the upper right corners offer a zoomed-in view of the footpoint switching at a different viewing angle, as evidenced by the lines switching colors.}
    \label{fig:sel5}
\end{figure*}

\begin{figure}
\begin{subfigure}{\textwidth}
  % \centering
  % include first image
  \hspace{0.02\linewidth}
  \includegraphics[width=0.405\linewidth]{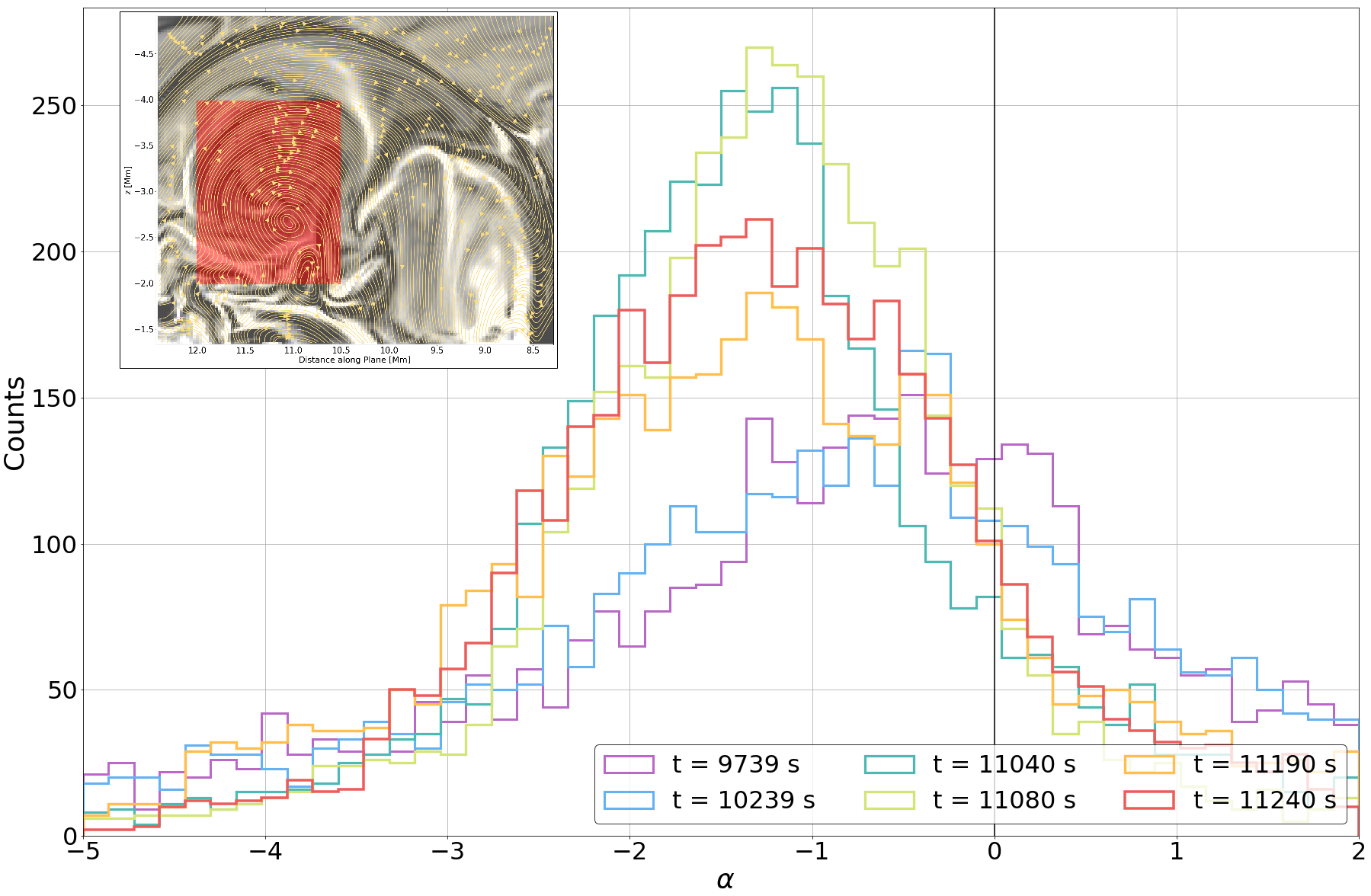}
\end{subfigure}
\begin{subfigure}{\textwidth}
  % \centering
  % include second image
  \includegraphics[width=0.5\linewidth]{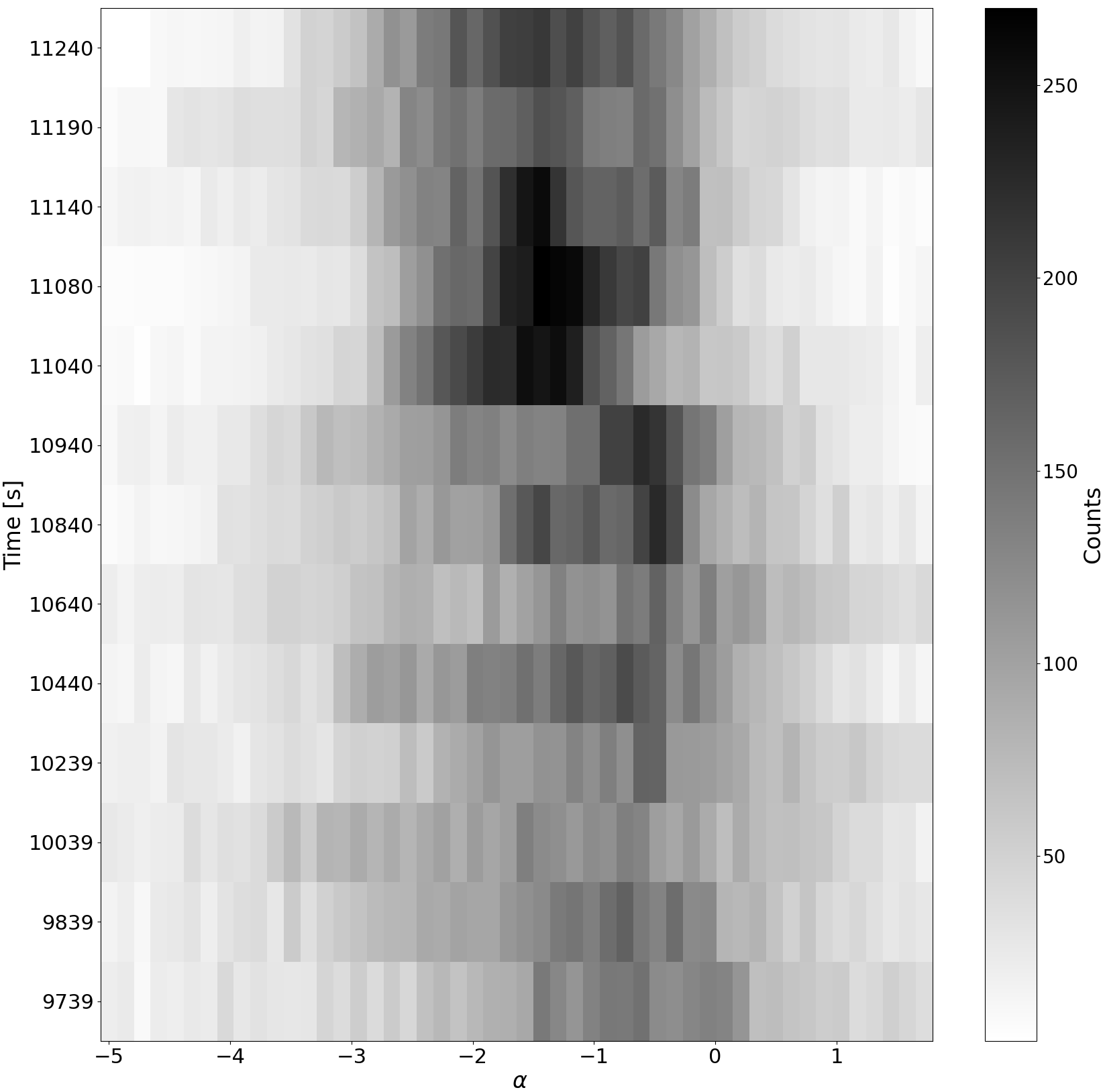}  
\end{subfigure}
\caption{Time series of $\alpha$ histograms corresponding to the same time stamps as in Figure \ref{fig:contours} (upper). The inset in the upper left corner shows the area within the flux rope center where these statistics were calculated. The lower panel shows a 2D histogram time series for several time stamps ranging from t = 9\,739 s until 11\,240 s.}
\label{fig:hist}
\end{figure}

We recall that the lower two panels of Figure \ref{fig:contours} demonstrate four small flux bundles (left) and one final helical state (right.) During that time span, we find several field line pairs that flicker between chromospheric footpoints. We consider upper-chromospheric flickering as indicators of footpoint switching because, in a fully stratified simulation such as this one, the environment in the chromosphere is so dynamic that the field integration in that region may not be as precise as is necessary to isolate individual line pairs. This is due to the existence of other QSLs down the line, which eventually diverge the magnetic field lines toward their respective photospheric roots. These lower-altitude reconnections play little role in flux rope formation in the atmosphere, so in order to effectively search for local atmospheric component reconnection, we make a chromospheric cut and search for pairs of coherent field lines that show signs of component reconnection by switching chromospheric footpoints.

To find reconnecting magnetic field line pairs, we searched for all available corks that pass through the vertical cut near the relevant current sheets, initially seeded by a random distribution of seeds within a cube encompassing that part of the rope. These seeds were integrated bidirectionally for 30 integration steps, then the closest corks to those line segments were used to search for magnetic field line pairs.

Figures \ref{fig:sel2} and \ref{fig:sel5} demonstrate two distinct field line pairs over these time stamps that show clear component reconnection behavior. In Figure \ref{fig:sel2}, the orange and red lines pass through the vertical cut just above a current sheet at t = 11\,190 s and t = 11\,200 s. A chromospheric sheet is given as a footpoint boundary, and between the left and right panels of Figure \ref{fig:sel2}, the red and orange lines appear to switch colors, indicating that they must have switched footpoints. The inset shows a clear picture of this, demonstrating that the red line becomes the orange line, and vice versa. We note that the two lines are traced by two consistent corks, meaning that under component reconnection, one cork traces one line at the first time stamp, but then the other reconnected line at the second time stamp.

Similar behavior is exhibited by the pair shown in Figure \ref{fig:sel5} between t = 11\,200 s (left) and t = 11\,210 s (right) where the green and purple lines switch behavior, again traced by two consistent corks. This behavior indicates footpoint switching, demonstrating that small-scale reconnection inferred in Section \ref{bundles} is indeed taking place via multiple component reconnections in field line pairs within the flux rope.

We have isolated two examples of this behavior, but finding such field line pairs depends on whether or not we have pairs of corks available near the relevant current sheets at those time stamps. Our success in finding these two examples, as well as the clear identification of current sheets inside the rope as especially seen in Figure \ref{fig:contours}, demonstrates that there are likely many more component reconnections occurring throughout the flux rope during the bulk mergers. This implies that the inverse cascade of helicity begins at these small scales; with component reconnection along magnetic field lines all the way up to the ordering of the wider flux rope. 

This apparent cascade of helicity from small, component reconnection scales to the larger scale of the flux rope should then be accompanied by some degree of magnetohydrodynamic relaxation toward a force-free field. With that, the question is whether or not the field approaches a linear (that is, constant $\alpha$) force-free field. To confirm that there is indeed a shift and settling in $\alpha$ values during flux rope formation, Figure \ref{fig:hist} shows 1D (upper) and 2D (lower) histogram time series of $\alpha$ within the rope center. The 1D histogram time series illustrates six different time stamps (the same as in Figure \ref{fig:contours}) which belong to three separate regimes: the first two selected time stamps occur before flux rope formation, the second two time stamps occur when the flux rope tightens, and the last two time stamps occurs where major reconnection has not yet occurred, but upper field lines are just beginning to peel back. 

These three regimes represent different distributions of $\alpha$; before flux rope formation, we see a broader distribution with higher counts at slightly smaller structures (that is, larger values of $|\alpha|$). During flux rope formation, we see a sharper distribution centered at roughly -1.3 Mm$^{-1}$. During the stages before major reconnection, we see the same peak value but a slightly broader distribution. By visual inspection of the twist per unit length in Figures \ref{fig:flux_compare} and \ref{fig:bigrope}, we see a left-handed, $\approx$ 10 Mm rope with one turn. With that, an $\alpha$ value of -1.3 Mm$^{-1}$ is reasonable and expected as per Equation \ref{eq:2}. 

The systematic shift in distributions over time is also seen in the lower panel of Figure \ref{fig:hist}, taking into account several more time stamps than only those given in Figure \ref{fig:contours}. Here, it is even more clear that the distributions begin relatively featureless, then tighten around -1.3 Mm$^{-1}$ around t = 11\,040 s, then broaden again slightly around t = 11\,190 s. A further analysis of this flux rope relaxation, and why we do not expect to see total relaxation, is given in the next section.

\section{Discussion}
\subsection{Comparisons to earlier studies}
While this study is comparable to several recent studies on inverse helicity cascade, it is also topologically comparable to early studies on flux rope formation. For example, the formation of a flux rope confined by a magnetic arcade has been analyzed in \citet{1999ApJ...515L..81A} and \citet{1999ApJ...518L..57A}. These letters discuss the mechanisms behind flux rope formation, and how a given topology can end up in a relaxed state that includes a flux rope and magnetic arcade. We do not aim to provide a solution for prominence support as they do in these studies, but the fundamental topological results are similar. We see the relaxation of the magnetic field into a flux rope and magnetic arcade \citep[as discussed in][]{2022A&A...668A.177R} meaning that our stratified and dynamic simulation relaxes (as much as it can relax) to a similar configuration as in unstratified simulations.

Since these earlier studies, significant work has been done to determine the details of flux rope formation in terms of relaxation and reconnection. As the simulation progresses, we see several interactions between smaller-scale twisting ropes which result in a larger-scale distribution of twist. This has been seen in plane-parallel simulations, for example, in \citet{Milano_1999,2007A&A...473..615W,2015ApJ...805...61Z} and \citet{2019ApJ...883..148R}. In such setups, systems of corotating flux tubes interact with each other until their magnetic field lines reconnect, and eventually the overall helicity of the system changes from small-scale twist to large-scale twist. This, as in our study, is a result of small reconnection events between individual field lines as the flux bundles rotate. In setups where the flux tubes are counterrotating, inverse helicity cascade cannot be reproduced because the conditions for reconnection are not met. We see that the same physical rules apply in our simulation; small-scale reconnection occurs where conditions are favorable, and does not occur where conditions are unfavorable.  

Our work is also complimentary to a similar study done by \citet{2017ApJ...835...85K}, wherein systems of flux tubes injected varying helicity into the corona via rotating drivers. This resulted in an inverse condensation of helicity and the self-organization of coronal structures. One of their conclusions was that, as long as the photospheric drivers have a helicity preference, helicity injection results in large-scale structuring of coronal features.

The simulations mentioned above, though less complex than \textit{Bifrost}, are complimentary to our simulation in that the fundamental physics represented in their work is consistent with ours: we also see the merging of small-scale twisted ropes into an overall large-scale twist just before the onset of our largest reconnection event. This coalescence follows small-scale reconnections within the overall flux rope. These reconnections are evidenced by the existence of current sheets within the flux rope itself as well as changes in connectivity between the corona and the photosphere. This footpoint flickering indicates that reconnection is taking place between lines that pass through consistent Lagrangian markers, and that the lines likely meet different QSL systems on their way to the photosphere. 

The fact that we see, to some degree, consistent physics between idealized setups and our simulation shows that these processes can be found and explored in fully stratified simulations. Simulations like ours are not too complex to isolate, for example, individual component reconnection events. The theory developed using idealized simulations serves as a guiding light for the analysis of more complex simulations, allowing us to further explore the effects of well-developed ideas like the inverse cascade of helicity. Of course, there are still some differences between idealized simulations and our simulation; the nature of our simulation allows us to explore processes that are difficult or impossible to model in idealized simulations. For example, we are able to simulate an environment that tends toward Taylor relaxation but cannot fully relax, as we discuss in the next subsection. 

\subsection{Incomplete Taylor relaxation in a dynamic simulation}
In the preceding sections, we have discussed the formation of a weakly twisted flux rope in our 3D, fully stratified, self-consistently driven MHD model using \textit{Bifrost}. We have demonstrated that the flux rope forms spontaneously from tangled coronal field lines via small-scale reconnection, beginning from component reconnection on the smallest observable scale, and condensing helicity from these small scales to the wider scale of the rope itself. Multiple reconnections are required to build such a well-defined and coherent flux rope, which is what we see here. We note that the initially disorganized field lines can be regarded as evidence of turbulent flows, and the presence of thin current sheets is evidence of the forward cascade of energy to smaller scales \citep[e.g.,][]{2019ApJ...883..148R,10.1093/mnras/stac3188}. With that, we see that the forward cascade of energy works in tandem with the inverse cascade of helicity. Then, according to Taylor's theory, the system would tend toward a relaxed state in which the entire computational domain approaches one value of $\alpha$, but this has been shown to be unrealistic for coronal magnetic fields. 

\citet{1999PPCF...41A.779A} demonstrates that the final relaxed state of their flux configuration after kinking and reconnecting with surrounding fields is two flux ropes; certainly not a linear force-free field. \citet{1999GMS...111..187A} argue that the relaxation of the global corona to a linear force-free field is never observed due to magnetic reconnection happening at only specified regions, effectively prohibiting the Sun from fully relaxing according to Taylor's theory. The simulations mentioned in these two studies, however, do not include many small-scale current sheets; only large-scale ones associated with larger-scale reconnection events. Therefore, no intermittent current sheets exist within the flux systems in question. With that, there would have been no basis for a Taylor relaxation to a linear force-free field, as their large-scale current sheets do not allow for reconnection to occur everywhere in the box.

Our simulation, as we have especially seen in Figure \ref{fig:contours}, is replete with small-scale and intermittent current sheets. These current sheets are reminiscent of small-scale reconnection events, and over time, the cascade of helicity from very small scales along field lines to the larger scale of the flux system. Figures \ref{fig:bigrope} and \ref{fig:contours} demonstrate qualitatively that we do see systematic smoothing of the small-scale current sheets to a larger-scale structure. However, as Figure \ref{fig:hist} shows quantitatively, we do not see complete Taylor relaxation to one value of $\alpha$ even when our small-scale conditions should encourage the simulation to attempt it. 

The first reason for this is that we have previously observed this flux system to undergo major reconnection with an arcade and overlying horizontal field, as discussed in \citet{2022A&A...668A.177R}. This reconnection, already beginning for some lines as seen in the later stages of Figures \ref{fig:bigrope}, \ref{fig:contours}, and \ref{fig:hist}, eventually rips the flux rope apart and generates energy on the scale of a nanoflare. Under such conditions, no further relaxation is possible.

Second, there exist QSLs and thin current sheets nearly everywhere in this simulation. The thinnest ones are, of course, involved in the eventual major reconnection event; however, we see current sheets within the flux rope system even when it is well into its relaxation process (see Figure \ref{fig:contours}). In a fully stratified simulation with self-consistent convective driving, the flux system is always supplied with new electric currents from the drivers and reconnection on small scales is essentially always taking place somewhere in the box. With such consistent driving, it is not possible for the flux system to fully relax according to Taylor's theory.

It is quite remarkable that under these dynamic conditions, we still see a tendency for the flux rope center to gravitate toward an expected $\alpha$ value of -1.3 Mm$^{-1}$ as shown in Figure \ref{fig:hist} and expected from Equation \ref{eq:2} as well as Figure \ref{fig:bigrope}. With perfect Taylor relaxation of the flux system, we would expect every gridpoint in the system to settle to exactly that value; but even with our more realistic driving and stratification, we still see signs of relaxation in the statistics. This is our final piece of compelling evidence to suggest that this flux rope, forming from disordered coronal field lines, finds order via the inverse cascade of helicity and incomplete Taylor relaxation from the scale of field line pairs to the scale of the rope itself.

\section{Conclusions}
We used a fully stratified, self-consistently driven \textit{Bifrost} simulation to explore spontaneous flux rope formation in the corona, and determined that the flux rope self-orders via the inverse cascade of helicity and incomplete Taylor relaxation. Our simulation provides a case study for the formation of flux systems which, under the right conditions, may eventually contribute to atmospheric heating via large-scale reconnection. Our conclusions are summarized as follows:

\begin{itemize}
    \item We follow the gradual buildup of a flux rope via multiple small-scale reconnections, before the flux system itself reconnects with an overlying field. This buildup is demonstrably a result of component reconnection between individual lines and small-scale reconnection between small flux bundles within the overall flux rope.
    \item We developed a new method for following component reconnection, as described in Section \ref{component}. Considering the network of complex photospheric footpoints in this simulation, in combination with many QSLs in the lower chromosphere, it was necessary to look for component reconnection via switching of chromospheric footpoints. Using this method, we isolated two line pairs that undergo component reconnection within the flux rope.
    \item We show that the flux rope attempts to relax to an $\alpha$ value of -1.3 Mm$^{-1}$, consistent with an $\approx$10 Mm left-handed rope with one turn, which is consistent with the overall twist of the representative field lines shown in Figure \ref{fig:bigrope}. This demonstrates that the inverse cascade of helicity also tends toward an incomplete Taylor relaxation.
    \item We present a yet overlooked mechanism for low-lying flux rope formation in the quiet Sun in addition to traditional flux emergence, flux cancellation, and tether-cutting reconnection.
    \item We present this case study as an example of how the self-ordering and relaxation of low-lying loops can contribute to low-lying nanoflares, which themselves contribute to coronal heating in the quiet Sun.
\end{itemize}

In summary, we conclude that component reconnection along individual magnetic field line pairs contributes to the formation of small flux bundles, which eventually coalesce into a larger-scale helical flux rope. This self-ordering is an example of the inverse cascade of helicity in a fully stratified simulation. An unprecedented finding is that the inverse cascade of helicity, in this case, also corresponds to incomplete Taylor relaxation in the flux rope center. It does not fully relax, in part, because the flux rope meets an overlying horizontal field and undergoes major reconnection instead. This is an interesting result itself: the self-ordering and incomplete Taylor relaxation ultimately cause the flux rope to build up enough magnetic energy to power the nanoflare-scale reconnection.

In \citet{2022A&A...668A.177R}, we analyzed a major reconnection event between the aforementioned flux rope, a magnetic arcade, and an overlying coronal field. We determined that the combined energy of that event was on the order of $5.4 \times 10^{17}$ J ($5.4 \times 10^{24}$ ergs), consistent with the nanoflare regime. We now bolster our previous study with more information about the formation of the flux rope, as its self-consistent and spontaneous formation is a consequence of convection-driven footpoint motion as well as small-scale reconnection within the rope. Equipped with information about the magnetic geometry of our simulated event, its energy content, and now the formation of a relevant flux rope under small-scale reconnection, we can now provide new insights into the mechanisms that can power nanoflares in the quiet Sun. 

\begin{acknowledgements}
      This research was supported by the Research Council of Norway through its Centres of Excellence scheme, project number 262622, and through grants of computing time from the Programme for Supercomputing. We acknowledge funding support by the European Research Council under ERC Syngergy grant agreement No. 810218 (Whole Sun). G.A. acknowledges financial support from the French national space agency (CNES), as well as from the Programme National Soleil Terre (PNST) of the CNRS/INSU also cofunded by CNES and CEA. R.R. acknowledges the \textit{Bifrost} developers at RoCS and thanks them for their guidance and support.
\end{acknowledgements}
     
\bibliography{biblio}

\begin{thebibliography}{59}
\expandafter\ifx\csname natexlab\endcsname\relax\def\natexlab#1{#1}\fi

\bibitem[{{Amari} \& {Luciani}(1999)}]{1999ApJ...515L..81A}
{Amari}, T. \& {Luciani}, J.~F. 1999, \apjl, 515, L81

\bibitem[{{Amari} {et~al.}(1999{\natexlab{a}}){Amari}, {Luciani}, \&
  {Mikic}}]{1999PPCF...41A.779A}
{Amari}, T., {Luciani}, J.~F., \& {Mikic}, Z. 1999{\natexlab{a}}, Plasma
  Physics and Controlled Fusion, 41, A779

\bibitem[{{Amari} {et~al.}(1999{\natexlab{b}}){Amari}, {Luciani}, {Mikic}, \&
  {Linker}}]{1999ApJ...518L..57A}
{Amari}, T., {Luciani}, J.~F., {Mikic}, Z., \& {Linker}, J. 1999{\natexlab{b}},
  \apjl, 518, L57

\bibitem[{{Antiochos} \& {DeVore}(1999)}]{1999GMS...111..187A}
{Antiochos}, S.~K. \& {DeVore}, C.~R. 1999, Geophysical Monograph Series, 111,
  187

\bibitem[{{Archontis} {et~al.}(2004){Archontis}, {Moreno-Insertis},
  {Galsgaard}, {Hood}, \& {O'Shea}}]{2004A&A...426.1047A}
{Archontis}, V., {Moreno-Insertis}, F., {Galsgaard}, K., {Hood}, A., \&
  {O'Shea}, E. 2004, \aap, 426, 1047

\bibitem[{{Aulanier}(2014)}]{2014IAUS..300..184A}
{Aulanier}, G. 2014, in Nature of Prominences and their Role in Space Weather,
  ed. B.~{Schmieder}, J.-M. {Malherbe}, \& S.~T. {Wu}, Vol. 300, 184--196

\bibitem[{{Aulanier} {et~al.}(2010){Aulanier}, {T{\"o}r{\"o}k}, {D{\'e}moulin},
  \& {DeLuca}}]{2010ApJ...708..314A}
{Aulanier}, G., {T{\"o}r{\"o}k}, T., {D{\'e}moulin}, P., \& {DeLuca}, E.~E.
  2010, \apj, 708, 314

\bibitem[{{Bakke} {et~al.}(2018){Bakke}, {Frogner}, \&
  {Gudiksen}}]{2018A&A...620L...5B}
{Bakke}, H., {Frogner}, L., \& {Gudiksen}, B.~V. 2018, \aap, 620, L5

\bibitem[{{Bellot Rubio} \& {Orozco Su{\'a}rez}(2019)}]{2019LRSP...16....1B}
{Bellot Rubio}, L. \& {Orozco Su{\'a}rez}, D. 2019, Living Reviews in Solar
  Physics, 16, 1

\bibitem[{{Berger}(1999)}]{1999PPCF...41B.167B}
{Berger}, M.~A. 1999, Plasma Physics and Controlled Fusion, 41, B167

\bibitem[{{Berger} \& {Prior}(2006)}]{2006JPhA...39.8321B}
{Berger}, M.~A. \& {Prior}, C. 2006, Journal of Physics A Mathematical General,
  39, 8321

\bibitem[{{Carlsson} \& {Leenaarts}(2012)}]{2012A&A...539A..39C}
{Carlsson}, M. \& {Leenaarts}, J. 2012, \aap, 539, A39

\bibitem[{{Cheng} {et~al.}(2014){Cheng}, {Ding}, {Zhang}, {Srivastava}, {Guo},
  {Chen}, \& {Sun}}]{2014ApJ...789L..35C}
{Cheng}, X., {Ding}, M.~D., {Zhang}, J., {et~al.} 2014, \apjl, 789, L35

\bibitem[{{Cheng} {et~al.}(2011){Cheng}, {Zhang}, {Liu}, \&
  {Ding}}]{2011ApJ...732L..25C}
{Cheng}, X., {Zhang}, J., {Liu}, Y., \& {Ding}, M.~D. 2011, \apjl, 732, L25

\bibitem[{{Cowley}(1976)}]{1976JGR....81.3455C}
{Cowley}, S.~W.~H. 1976, \jgr, 81, 3455

\bibitem[{{Demoulin} {et~al.}(1996){Demoulin}, {Henoux}, {Priest}, \&
  {Mandrini}}]{1996A&A...308..643D}
{Demoulin}, P., {Henoux}, J.~C., {Priest}, E.~R., \& {Mandrini}, C.~H. 1996,
  \aap, 308, 643

\bibitem[{{Druett} {et~al.}(2022){Druett}, {Leenaarts}, {Carlsson}, \&
  {Szydlarski}}]{2022A&A...665A...6D}
{Druett}, M.~K., {Leenaarts}, J., {Carlsson}, M., \& {Szydlarski}, M. 2022,
  \aap, 665, A6

\bibitem[{{Fan}(2009)}]{2009ApJ...697.1529F}
{Fan}, Y. 2009, \apj, 697, 1529

\bibitem[{{Frisch} {et~al.}(1975){Frisch}, {Pouquet}, {Leorat}, \&
  {Mazure}}]{1975JFM....68..769F}
{Frisch}, U., {Pouquet}, A., {Leorat}, J., \& {Mazure}, A. 1975, Journal of
  Fluid Mechanics, 68, 769

\bibitem[{{Gudiksen} {et~al.}(2011){Gudiksen}, {Carlsson}, {Hansteen}, {Hayek},
  {Leenaarts}, \& {Mart{\'\i}nez-Sykora}}]{2011A&A...531A.154G}
{Gudiksen}, B.~V., {Carlsson}, M., {Hansteen}, V.~H., {et~al.} 2011, \aap, 531,
  A154

\bibitem[{{Hannah} {et~al.}(2011){Hannah}, {Hudson}, {Battaglia}, {Christe},
  {Ka{\v{s}}parov{\'a}}, {Krucker}, {Kundu}, \&
  {Veronig}}]{2011SSRv..159..263H}
{Hannah}, I.~G., {Hudson}, H.~S., {Battaglia}, M., {et~al.} 2011, \ssr, 159,
  263

\bibitem[{{Hayek} {et~al.}(2010){Hayek}, {Asplund}, {Carlsson}, {Trampedach},
  {Collet}, {Gudiksen}, {Hansteen}, \& {Leenaarts}}]{2010A&A...517A..49H}
{Hayek}, W., {Asplund}, M., {Carlsson}, M., {et~al.} 2010, \aap, 517, A49

\bibitem[{Hyman(1979)}]{hyman}
Hyman, J. 1979, Advances in Computer Methods for Partial Differential Equations
  (Vichnevetsky, R. and Stepleman, R.S. and International Association for
  Mathematics and Computers in Simulation)

\bibitem[{{Jiang} {et~al.}(2019){Jiang}, {Duan}, {Feng}, {Zou}, {Zuo}, \&
  {Wang}}]{2019FrASS...6...63J}
{Jiang}, C., {Duan}, A., {Feng}, X., {et~al.} 2019, Frontiers in Astronomy and
  Space Sciences, 6, 63

\bibitem[{{Keppens} {et~al.}(2014){Keppens}, {Porth}, \&
  {Xia}}]{2014ApJ...795...77K}
{Keppens}, R., {Porth}, O., \& {Xia}, C. 2014, \apj, 795, 77

\bibitem[{{Knizhnik} {et~al.}(2017){Knizhnik}, {Antiochos}, \&
  {DeVore}}]{2017ApJ...835...85K}
{Knizhnik}, K.~J., {Antiochos}, S.~K., \& {DeVore}, C.~R. 2017, \apj, 835, 85

\bibitem[{{Li} {et~al.}(2019){Li}, {Jaroszynski}, {Pearse}, {Orf}, \&
  {Clyne}}]{2019Atmos..10..488L}
{Li}, S., {Jaroszynski}, S., {Pearse}, S., {Orf}, L., \& {Clyne}, J. 2019,
  Atmosphere, 10, 488

\bibitem[{{Liu} {et~al.}(2016){Liu}, {Kliem}, {Titov}, {Chen}, {Wang}, {Wang},
  {Liu}, {Xu}, \& {Wiegelmann}}]{2016ApJ...818..148L}
{Liu}, R., {Kliem}, B., {Titov}, V.~S., {et~al.} 2016, \apj, 818, 148

\bibitem[{{Makwana} {et~al.}(2018){Makwana}, {Keppens}, \&
  {Lapenta}}]{2018PhPl...25h2904M}
{Makwana}, K.~D., {Keppens}, R., \& {Lapenta}, G. 2018, Physics of Plasmas, 25,
  082904

\bibitem[{Milano {et~al.}(1999)Milano, Dmitruk, Mandrini, Gómez, \&
  Démoulin}]{Milano_1999}
Milano, L.~J., Dmitruk, P., Mandrini, C.~H., Gómez, D.~O., \& Démoulin, P.
  1999, The Astrophysical Journal, 521, 889

\bibitem[{{Moore} \& {Roumeliotis}(1992)}]{1992LNP...399...69M}
{Moore}, R.~L. \& {Roumeliotis}, G. 1992, in IAU Colloq. 133: Eruptive Solar
  Flares, ed. Z.~{Svestka}, B.~V. {Jackson}, \& M.~E. {Machado}, Vol. 399, 69

\bibitem[{{Moore} {et~al.}(2002){Moore}, {Fok}, \&
  {Chandler}}]{2002JGRA..107.1332M}
{Moore}, T.~E., {Fok}, M.~C., \& {Chandler}, M.~O. 2002, Journal of Geophysical
  Research (Space Physics), 107, 1332

\bibitem[{{Nordlund}(1982)}]{1982A&A...107....1N}
{Nordlund}, A. 1982, \aap, 107, 1

\bibitem[{{Nordlund} \& {Galsgaard}(1995)}]{1995NordlundGalsgaard}
{Nordlund}, {\AA}. \& {Galsgaard}, K. 1995, A 3D MHD Code for Parallel
  Computers

\bibitem[{Pariat(2020)}]{Pariat2020}
Pariat, {\'E}. 2020, Using Magnetic Helicity, Topology, and Geometry to
  Investigate Complex Magnetic Fields, ed. D.~MacTaggart \& A.~Hillier (Cham:
  Springer International Publishing), 145--175

\bibitem[{{Parker}(1988)}]{1988ApJ...330..474P}
{Parker}, E.~N. 1988, \apj, 330, 474

\bibitem[{{Pouquet} {et~al.}(2019){Pouquet}, {Rosenberg}, {Stawarz}, \&
  {Marino}}]{2019E&SS....6..351P}
{Pouquet}, A., {Rosenberg}, D., {Stawarz}, J.~E., \& {Marino}, R. 2019, Earth
  and Space Science, 6, 351

\bibitem[{{Prior} \& {Yeates}(2016{\natexlab{a}})}]{2016A&A...587A.125P}
{Prior}, C. \& {Yeates}, A.~R. 2016{\natexlab{a}}, \aap, 587, A125

\bibitem[{{Prior} \& {Yeates}(2016{\natexlab{b}})}]{2016A&A...591A..16P}
{Prior}, C. \& {Yeates}, A.~R. 2016{\natexlab{b}}, \aap, 591, A16

\bibitem[{{Rappazzo} {et~al.}(2019){Rappazzo}, {Velli}, {Dahlburg}, \&
  {Einaudi}}]{2019ApJ...883..148R}
{Rappazzo}, A.~F., {Velli}, M., {Dahlburg}, R.~B., \& {Einaudi}, G. 2019, \apj,
  883, 148

\bibitem[{Reid {et~al.}(2022)Reid, Threlfall, \& Hood}]{10.1093/mnras/stac3188}
Reid, J., Threlfall, J., \& Hood, A.~W. 2022, Monthly Notices of the Royal
  Astronomical Society, 518, 1584

\bibitem[{{Rempel}(2017)}]{2017ApJ...834...10R}
{Rempel}, M. 2017, \apj, 834, 10

\bibitem[{{Richard} {et~al.}(1990){Richard}, {Sydora}, \&
  {Ashour-Abdalla}}]{1990PhFlB...2..488R}
{Richard}, R.~L., {Sydora}, R.~D., \& {Ashour-Abdalla}, M. 1990, Physics of
  Fluids B, 2, 488

\bibitem[{{Robinson} {et~al.}(2022){Robinson}, {Carlsson}, \&
  {Aulanier}}]{2022A&A...668A.177R}
{Robinson}, R.~A., {Carlsson}, M., \& {Aulanier}, G. 2022, \aap, 668, A177

\bibitem[{{Skartlien}(2000)}]{2000ApJ...536..465S}
{Skartlien}, R. 2000, \apj, 536, 465

\bibitem[{{Song} {et~al.}(2015){Song}, {Chen}, {Zhang}, {Cheng}, {Wang}, {Hu},
  {Li}, \& {Wang}}]{2015ApJ...808L..15S}
{Song}, H.~Q., {Chen}, Y., {Zhang}, J., {et~al.} 2015, \apjl, 808, L15

\bibitem[{{Taylor}(1974)}]{1974PhRvL..33.1139T}
{Taylor}, J.~B. 1974, \prl, 33, 1139

\bibitem[{{Testa} {et~al.}(2014){Testa}, {De Pontieu}, {Allred}, {Carlsson},
  {Reale}, {Daw}, {Hansteen}, {Martinez-Sykora}, {Liu}, {DeLuca}, {Golub},
  {McKillop}, {Reeves}, {Saar}, {Tian}, {Lemen}, {Title}, {Boerner},
  {Hurlburt}, {Tarbell}, {Wuelser}, {Kleint}, {Kankelborg}, \&
  {Jaeggli}}]{2014Sci...346B.315T}
{Testa}, P., {De Pontieu}, B., {Allred}, J., {et~al.} 2014, Science, 346,
  1255724

\bibitem[{{Testa} {et~al.}(2013){Testa}, {De Pontieu}, {Mart{\'\i}nez-Sykora},
  {DeLuca}, {Hansteen}, {Cirtain}, {Winebarger}, {Golub}, {Kobayashi},
  {Korreck}, {Kuzin}, {Walsh}, {DeForest}, {Title}, \&
  {Weber}}]{2013ApJ...770L...1T}
{Testa}, P., {De Pontieu}, B., {Mart{\'\i}nez-Sykora}, J., {et~al.} 2013,
  \apjl, 770, L1

\bibitem[{{Titov} \& {D{\'e}moulin}(1999)}]{1999A&A...351..707T}
{Titov}, V.~S. \& {D{\'e}moulin}, P. 1999, \aap, 351, 707

\bibitem[{{Trattner} {et~al.}(2007){Trattner}, {Mulcock}, {Petrinec}, \&
  {Fuselier}}]{2007JGRA..112.8210T}
{Trattner}, K.~J., {Mulcock}, J.~S., {Petrinec}, S.~M., \& {Fuselier}, S.~A.
  2007, Journal of Geophysical Research (Space Physics), 112, A08210

\bibitem[{{van Ballegooijen} \& {Martens}(1989)}]{1989ApJ...343..971V}
{van Ballegooijen}, A.~A. \& {Martens}, P.~C.~H. 1989, \apj, 343, 971

\bibitem[{{Visualization \& Analysis Systems Technologies}(2022)}]{vapor}
{Visualization \& Analysis Systems Technologies}. 2022, Visualization and
  Analysis Platform for Ocean, Atmosphere, and Solar Researchers (VAPOR version
  3.6.1)[Software], Boulder, CO: UCAR/NCAR - Computational and Information
  System Lab.

\bibitem[{{Wilmot-Smith} \& {De Moortel}(2007)}]{2007A&A...473..615W}
{Wilmot-Smith}, A.~L. \& {De Moortel}, I. 2007, \aap, 473, 615

\bibitem[{Yeates(2020)}]{Yeates2020}
Yeates, A.~R. 2020, Magnetohydrodynamic Relaxation Theory, ed. D.~MacTaggart \&
  A.~Hillier (Cham: Springer International Publishing), 117--143

\bibitem[{{Yeates} {et~al.}(2010){Yeates}, {Hornig}, \&
  {Wilmot-Smith}}]{2010PhRvL.105h5002Y}
{Yeates}, A.~R., {Hornig}, G., \& {Wilmot-Smith}, A.~L. 2010, \prl, 105, 085002

\bibitem[{{Zacharias} {et~al.}(2018){Zacharias}, {Hansteen}, {Leenaarts},
  {Carlsson}, \& {Gudiksen}}]{2018A&A...614A.110Z}
{Zacharias}, P., {Hansteen}, V.~H., {Leenaarts}, J., {Carlsson}, M., \&
  {Gudiksen}, B.~V. 2018, \aap, 614, A110

\bibitem[{{Zhao} {et~al.}(2015){Zhao}, {DeVore}, {Antiochos}, \&
  {Zurbuchen}}]{2015ApJ...805...61Z}
{Zhao}, L., {DeVore}, C.~R., {Antiochos}, S.~K., \& {Zurbuchen}, T.~H. 2015,
  \apj, 805, 61

\bibitem[{Zhukov(2002)}]{zhukov}
Zhukov, V. 2002, Plasma Physics Reports, 28, 411

\end{thebibliography}

\begin{appendix}
\section{Using the correct seeding method for finding relevant flux systems}
To figure out where any flux system comes from, we can trace the field by choosing seed coordinates that are either confined in a fixed volume throughout time, or move with the flow of the fluid. We can derive a more reliable time series from the latter, but for comparison, it is worth exploring the former to determine whether or not the flux system in question exists in that same space throughout the run. 

\begin{figure*}
\begin{subfigure}{.5\textwidth}
  \centering
  \begin{tikzpicture}
  \node[anchor=south west,inner sep=0] (image) at (0,0)
  {\includegraphics[width=.8\linewidth]{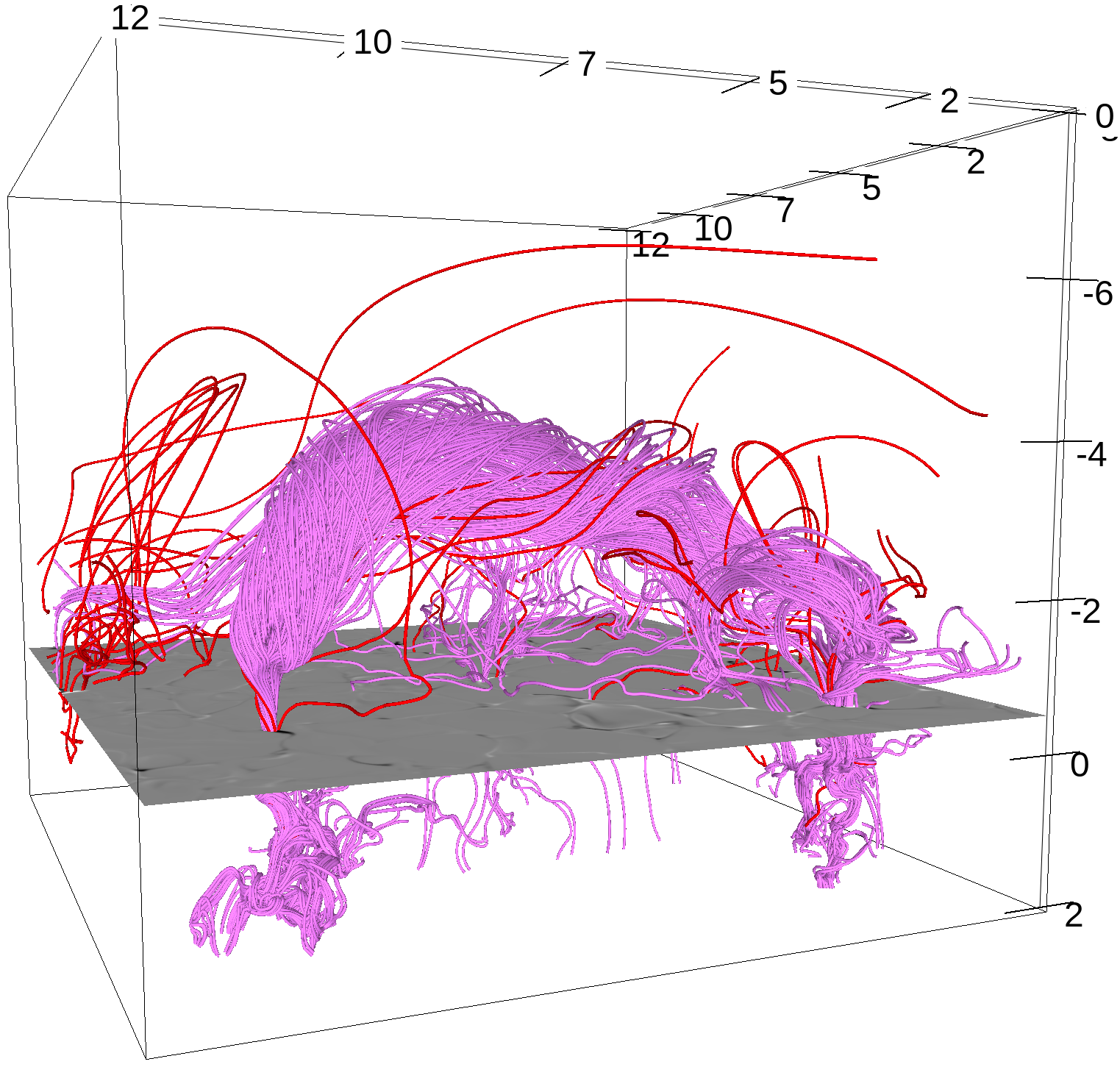}};
  \begin{scope}[x={(image.south east)},y={(image.north west)}]
    \node[black,fill=white,draw] at (0.2,0.85) {t = 9\,669 s};
  \end{scope}
  \end{tikzpicture}
  \subcaption[]{Seeds: random distribution at t = 9\,669 s}
\end{subfigure}
\begin{subfigure}{.5\textwidth}
  \centering
  \begin{tikzpicture}
  \node[anchor=south west,inner sep=0] (image) at (0,0)
  {\includegraphics[width=.8\linewidth]{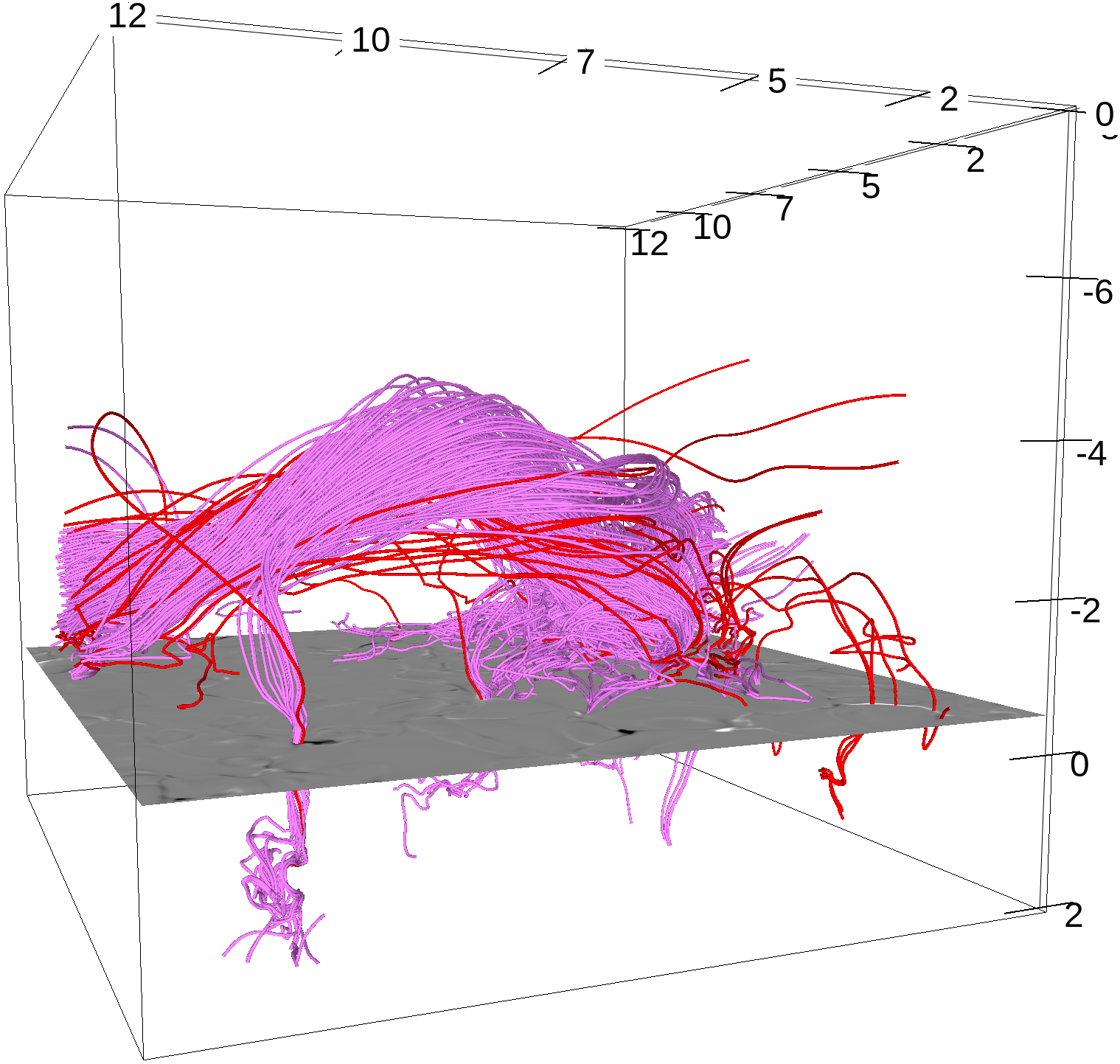}};
  \begin{scope}[x={(image.south east)},y={(image.north west)}]
    \node[black,fill=white,draw] at (0.2,0.85) {t = 11\,040 s};
  \end{scope}
  \end{tikzpicture}
  \subcaption[]{Seeds: random distribution at t = 11\,040 s} 
\end{subfigure}
\begin{subfigure}{.5\textwidth}
  \centering
  \begin{tikzpicture}
  \node[anchor=south west,inner sep=0] (image) at (0,0)
  {\includegraphics[width=.8\linewidth]{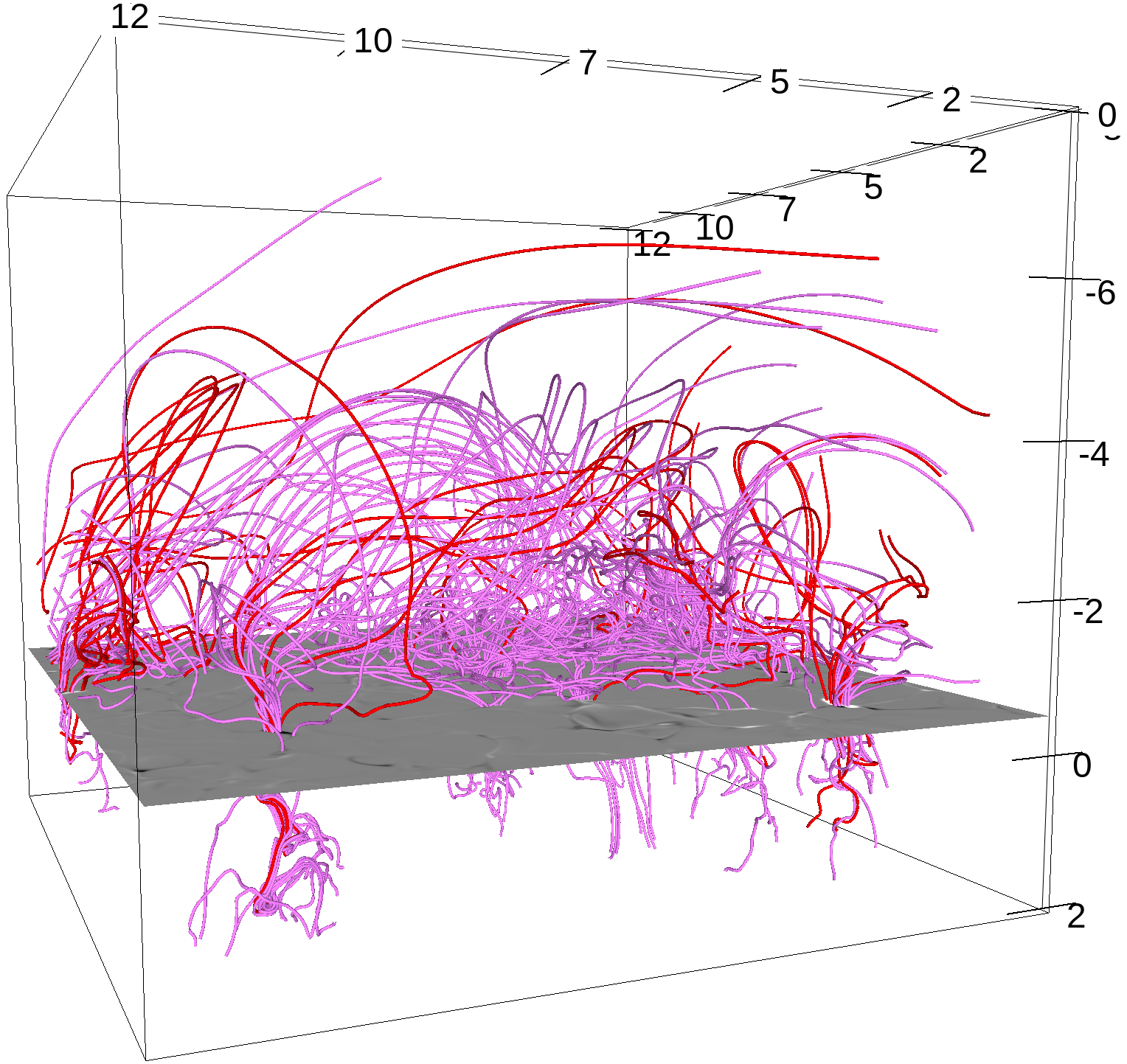}};
  \begin{scope}[x={(image.south east)},y={(image.north west)}]
    \node[black,fill=white,draw] at (0.2,0.85) {t = 9\,669 s};
  \end{scope}
  \end{tikzpicture}
  \subcaption[]{Seeds: Lagrangian markers selected at t = 11\,040 s}
\end{subfigure}
\begin{subfigure}{.5\textwidth}
  \centering
  \begin{tikzpicture}
  \node[anchor=south west,inner sep=0] (image) at (0,0)
  {\includegraphics[width=.8\linewidth]{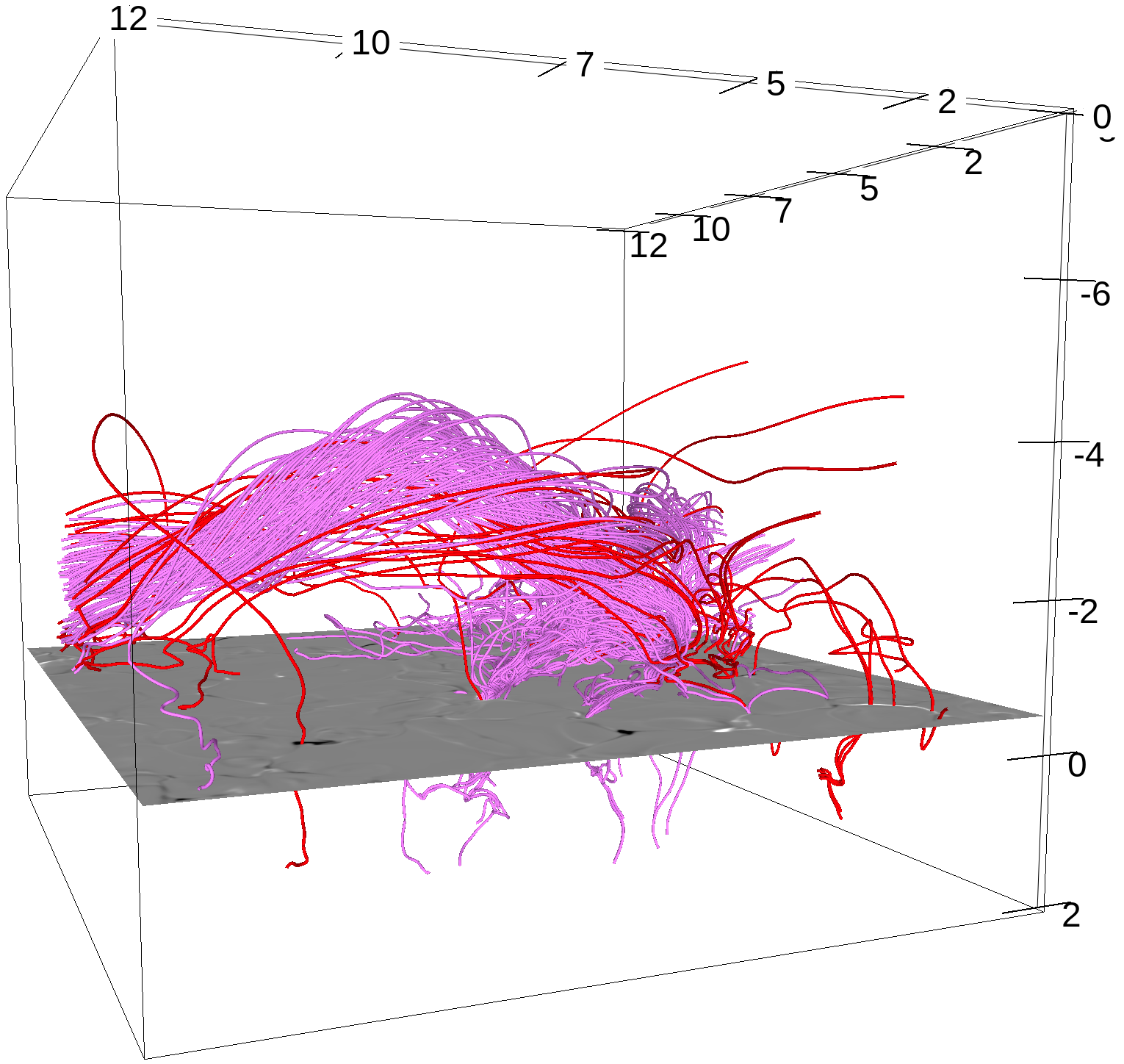}};
  \begin{scope}[x={(image.south east)},y={(image.north west)}]
    \node[black,fill=white,draw] at (0.2,0.85) {t = 11\,040 s};
  \end{scope}
  \end{tikzpicture}
  \subcaption[]{Seeds: Lagrangian markers selected at t = 11\,040 s} 
\end{subfigure}
\begin{subfigure}{.5\textwidth}
  \centering
  \begin{tikzpicture}
  \node[anchor=south west,inner sep=0] (image) at (0,0)
  {\includegraphics[width=.8\linewidth]{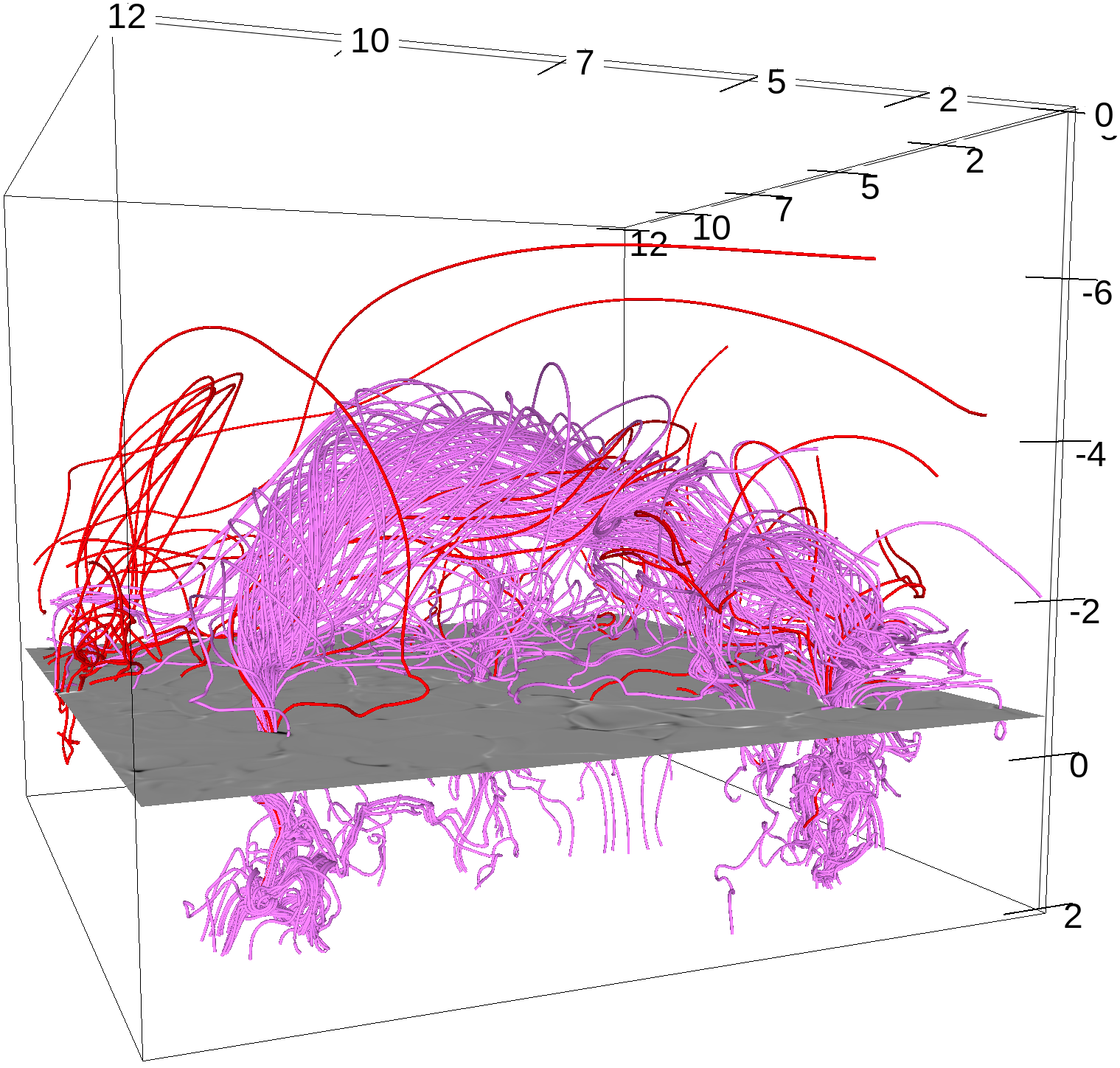}};
  \begin{scope}[x={(image.south east)},y={(image.north west)}]
    \node[black,fill=white,draw] at (0.2,0.85) {t = 9\,669 s};
  \end{scope}
  \end{tikzpicture}
  \subcaption[]{Seeds: Lagrangian markers selected at t = 9\,669 s}
\end{subfigure}
\begin{subfigure}{.5\textwidth}
  \centering
  \begin{tikzpicture}
  \node[anchor=south west,inner sep=0] (image) at (0,0)
  {\includegraphics[width=.8\linewidth]{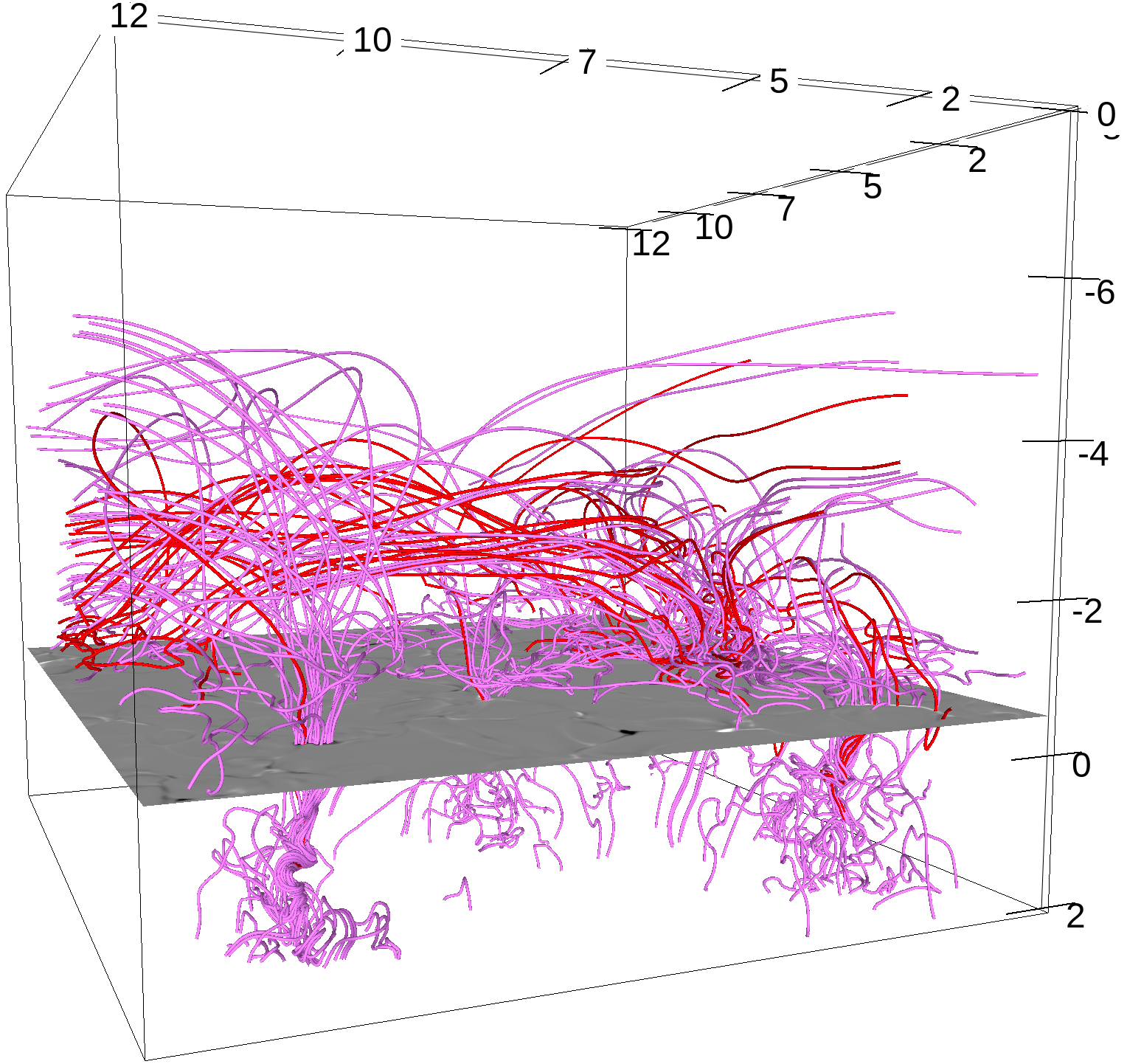}};
  \begin{scope}[x={(image.south east)},y={(image.north west)}]
    \node[black,fill=white,draw] at (0.2,0.85) {t = 11\,040 s};
  \end{scope}
  \end{tikzpicture}
  \subcaption[]{Seeds: Lagrangian markers selected at t = 9\,669 s} 
\end{subfigure}

\caption{Upper panels illustrate a horizontal flux system seeded consistently in space at t = 9\,669 s (left) and t = 11\,040 s (right). Red lines are seeded with Lagrangian markers as reported in \citet{2022A&A...668A.177R}. Center panels illustrate the time evolution of a horizontal flux system seeded by Lagrangian markers at t = 11\,040 s (right) and traced backward in time (left). Lower panels are the same as center panels, except seeded by Lagrangian markers beginning at t = 9\,669 s (left) and traced forward in time (right).}
\label{fig:seeding}
\end{figure*}
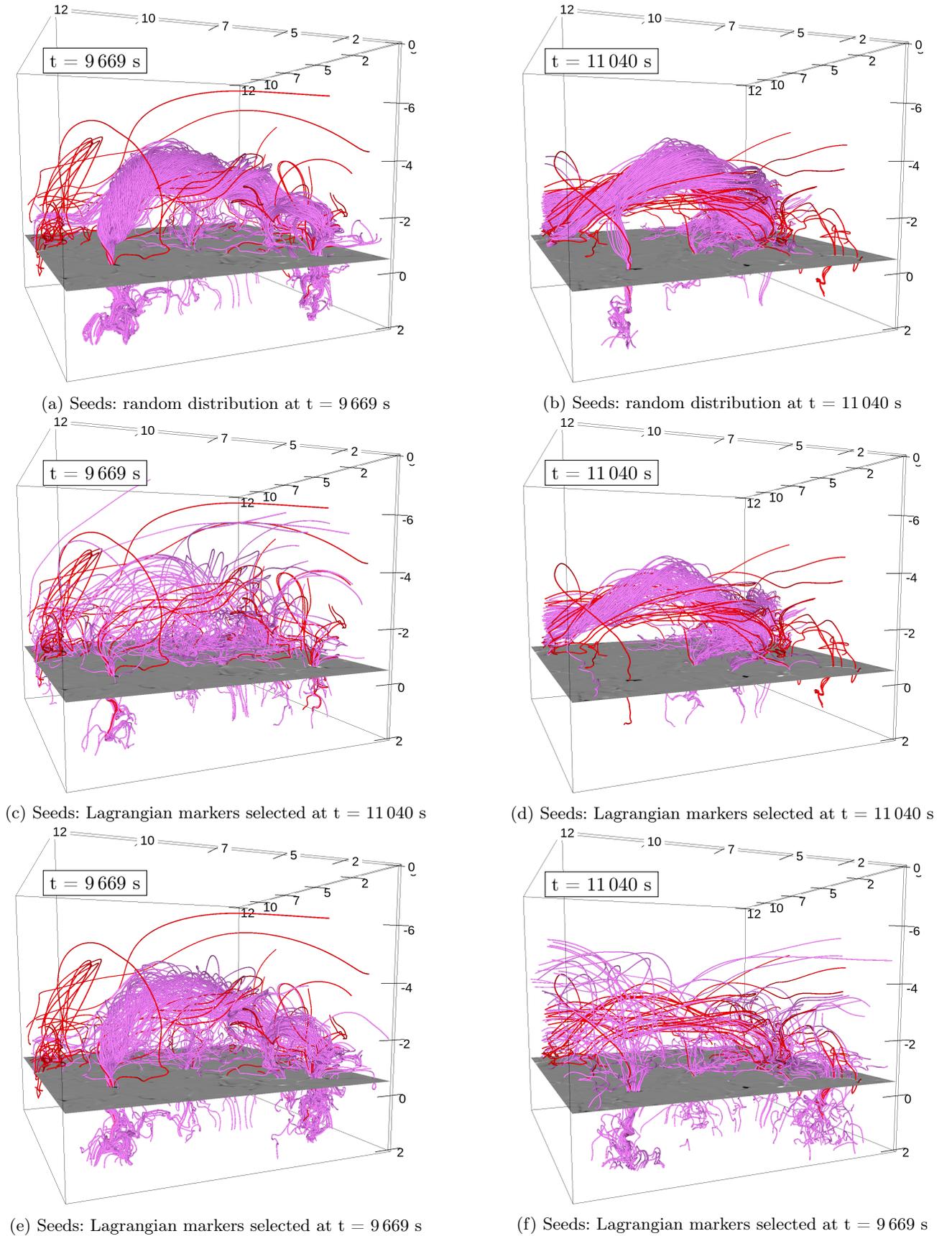

Figure \ref{fig:seeding} represents a comparison study of three seeding methods: one with random seeds through a consistent volume, and two with Lagrangian markers seeded at different time stamps. The results of these seeding methods are shown for two notable time stamps: t = 9\,669 s, which is the first time stamp where we have injected Lagrangian markers in the simulation, and t = 11\,040 s, where the horizontal flux rope has nearly formed but has not yet undergone major reconnection. For all panels in Figure \ref{fig:seeding}, the red lines represent the flux rope lines we presented in \citet{2022A&A...668A.177R}, which were seeded using specific Lagrangian markers. These red lines are added to emphasize that we are looking for and analyzing the same flux rope as in previous studies. 

The upper two panels of Figure \ref{fig:seeding} illustrate a horizontal flux system (pink lines) seeded randomly within a rectangular volume that encompasses the rope center. This seeding method is consistent in time, meaning that for all time stamps, the magnetic field is seeded by a random distribution of points within that volume. This does not necessarily trace consistent magnetic field lines, but rather traces whatever lines exist in that volume for each time stamp.

With this random seeding method, the selection has no prescribed bias nor preference for lines that had previously been selected. Because of that, the resulting field line tracing could be any random flux system that passes through that rectangular volume at the given time. The upper two panels of Figure \ref{fig:seeding} demonstrate that, with this seeding method, a flux system exists at t = 9\,669 s as well as at 11\,040 s; but they are not necessarily the same flux system. The flux system at t = 9\,669 s is firmly rooted in a strong negative photospheric footpoint, whereas the flux system at t = 11\,040 s is connected over the horizontal boundary on the left side, as we had seen in \citet{2022A&A...668A.177R}. Indeed, the red lines and pink lines are in much better agreement at t = 11\,040 s than at t = 9\,669 s.

By inspecting only the upper two panels of Figure \ref{fig:seeding}, it would seem as if the relevant flux system has always existed for this time span, and the most important event would be the change in connectivity from the strong negative polarity to the cross-boundary negative polarities on the other side. This would be a reasonable conclusion following the random seeding method, but it is a misleading one. 

If we instead employ a set of Lagrangian markers to trace the flux system, we are then able to trace consistent field lines rather than random lines seeded within a consistent volume. The flux system in question is nearly formed at t = 11\,040 s, so if the goal is to understand where those particular lines come from, then we must select Lagrangian markers closest to the flux system at that time and follow those same markers back in time to t = 9\,669 s. Alternatively, we could select Lagrangian markers closest to the flux system that exists at t = 9\,669 s and follow the markers forward in time to t = 11\,040 s. If the tracings agree both forward and backward in time, then the two flux systems are likely comprised of the same field lines. If not, then the lines cannot belong to the same flux system.

The two center panels of Figure \ref{fig:seeding} illustrate a flux system seeded by Lagrangian markers selected at t = 11\,040 s and traced backward in time to t = 9\,669 s. The two lower panels of Figure \ref{fig:seeding} also illustrate a flux system seeded by Lagrangian markers, but seeded at t = 9\,669 s and followed forward in time to t = 11\,040 s. At first glance, we see that these four panels do not agree with one another so we can immediately conclude that the two flux systems illustrated in the upper two panels of Figure \ref{fig:seeding} are \textit{not} comprised of the same magnetic field lines. 

Furthermore, the center two panels of Figure \ref{fig:seeding} illustrate a flux system that seems to form from a collection of incoherent lines. The pink lines and red lines are in good agreement in terms of their evolution from incoherence to relative coherence. The lower two panels, on the other hand, illustrate a flux system that begins as a relatively coherent (albeit fragmented) flux rope that falls apart by t = 11\,040 s. In this case, the red and pink lines do not agree well in terms of their coherence. Therefore, we conclude that the center two panels of Figure \ref{fig:seeding} best represent the flux system that had been presented in \citet{2022A&A...668A.177R}, which is also the same flux system that we further analyze in this work.

This comparison study has several implications. First, as we have already noted, the seemingly consistent flux systems illustrated in the upper two panels of Figure \ref{fig:seeding} are actually \textit{not} comprised of consistent magnetic field lines. We know this because of the two subsequent tests we ran using Lagrangian markers, illustrated in the center and lower panels of Figure \ref{fig:seeding}. This means that we must be very careful in choosing methods for tracing magnetic field lines, as Lagrangian markers are more likely to follow consistent magnetic field lines and are therefore more reliable for tracing the evolution of a given flux system.

Furthermore, it is necessary to have some \textit{a priori} knowledge of the relevant flux system in order to properly choose Lagrangian markers. In our case, we had already determined the approximate location of the flux system in question, as well as its approximate time of formation. Using that information, we could select Lagrangian markers associated with a given location and time in order to follow the system backwards in time, as in the center panels of Figure \ref{fig:seeding}. If we were to select Lagrangian markers too early and follow them forward in time instead, we would not be able to trace the flux system of interest as illustrated by the lower panels of Figure \ref{fig:seeding}. 

This exercise demonstrates why we must be cautious when choosing seeding methods, locations, and times in order to trace a flux system of interest. In our case, the red lines from previous analysis provided a baseline for the spatiotemporal information necessary for choosing initial Lagrangian markers. The relevant flux system is indeed the one given in the center panels of Figure \ref{fig:seeding}, and given the ability to use Lagrangian markers, we are confident that this flux system began incoherently and became coherent via small-scale reconnection processes. 

\label{appendix:a}

\end{appendix}

\end{document}